\documentclass[a4paper,fleqn,usenatbib]{mnras}

\usepackage{newtxtext,newtxmath}

\usepackage[T1]{fontenc}
\usepackage{ae,aecompl}

\usepackage{graphicx}	
\usepackage{amsmath}	
\usepackage{wasysym}

\def\mrpd{\hbox{mrad\,d$^{-1}$}}

\def\chisqr{\hbox{$\chi^2_{\rm r}$}}
\def\msun{\hbox{${\rm M}_{\odot}$}}
\def\mjup{\hbox{${\rm M}_{\jupiter}$}}

\def\mspy{\hbox{${\rm M}_{\odot}$\,yr$^{-1}$}}
\def\rsun{\hbox{${\rm R}_{\odot}$}}
\def\lsun{\hbox{${\rm L}_{\odot}$}}
\def\rcor{\hbox{$r_{\rm cor}$}}
\def\rmag{\hbox{$r_{\rm mag}$}}
\def\mstar{\hbox{$M_{\star}$}}
\def\rstar{\hbox{$R_{\star}$}}
\def\lstar{\hbox{$L_{\star}$}}
\def\teff{\hbox{$T_{\rm eff}$}}
\def\logg{\hbox{$\log g$}}

\def\kms{\hbox{km\,s$^{-1}$}}

\def\vsini{\hbox{$v \sin i$}}

\def\AV{\hbox{$A_{\rm V}$}}

\def\mic{\hbox{$\mu$m}}

\def\emr{}
\def\Bl{\hbox{$B_{\rm \ell}$}}
\def\Bd{\hbox{$B_{\rm d}$}}

\def\degr{\hbox{$^\circ$}}

\def\Mdot{\hbox{$\dot{M}$}}

\def\Prot{\hbox{$P_{\rm rot}$}}

\newcommand{\cai}{\hbox{Ca$\;${\sc i}}}
\newcommand{\tii}{\hbox{Ti$\;${\sc i}}}
\newcommand{\foxi}{\hbox{[O$\;${\sc i}]}}

\newcommand{\hei}{\hbox{He$\;${\sc i}}}
\newcommand{\hal}{\hbox{H${\alpha}$}}

\newcommand{\pab}{\hbox{Pa${\beta}$}}
\newcommand{\brg}{\hbox{Br${\gamma}$}}

\title[Spectropolarimetry \& velocimetry of PDS~70]{SPIRou observations of the young planet-hosting star PDS~70} 
\author[J.-F.~Donati et al.]{J.-F.~Donati$^{1}$\thanks{E-mail: jean-francois.donati@irap.omp.eu}, 
           P.I.~Cristofari$^{2}$, S.H.P.~Alencar$^{3}$, \'A.~K\'osp\'al$^{4,5}$, J.~Bouvier$^{6}$, C.~Moutou$^{1}$, 
\newauthor A.~Carmona$^{6}$, J.~Gregorio-Hetem$^{7}$, C.F.~Manara$^{8}$, E.~Artigau$^{9}$, R.~Doyon$^{9}$, M.~Takami$^{10}$,  
\newauthor H.~Shang$^{10}$, J.~Dias do Nascimento$^{11}$, F.~M\'enard$^{6}$, E.~Gaidos$^{12}$
and the SPIRou science team  
\vspace{1mm}
\\ 
$^1$ Universit\'e de Toulouse, CNRS, IRAP, 14 avenue Belin, 31400 Toulouse, France  \\ 
$^2$ Center for Astrophysics, Harvard \& Smithsonian, 60 Garden street, Cambridge, MA 02138, United States \\ 
$^3$ Departamento de F\'{\i}sica -- ICEx -- UFMG, Av. Ant\^onio Carlos, 6627, 30270-901 Belo Horizonte, MG, Brazil \\ 
$^4$ Konkoly Observatory, HUN-REN Research Centre for Astronomy and Earth Sciences, Konkoly-Thege Mikl\'os \'ut 15-17, 1121 Budapest, Hungary \\
$^5$ Institute of Physics and Astronomy, ELTE E\"otv\"os Lor\'and University, P\'azm\'any P\'eter s\'et\'any 1/A, 1117 Budapest, Hungary \\
$^6$ Universit\'e Grenoble Alpes, CNRS, IPAG, 38000 Grenoble, France  \\
$^7$ Universidade de S\~ao Paulo, IAG, Rua do Mat\~ao 1226, 05508-090 S\~ao Paulo, SP, Brazil \\ 
$^8$ European Southern Observatory, Karl-Schwarzschild-Strasse 2, 85748 Garching bei M\"unchen, Germany \\ 
$^9$ Universit\'e de Montr\'eal, D\'epartement de Physique, IREX, Montr\'eal, QC H3C 3J7, Canada  \\ 
$^{10}$ Institute of Astronomy and Astrophysics, Academia Sinica, Roosevelt Rd, Taipei 10617, Taiwan \\
$^{11}$ Dep. de F\'{\i}sica, UFRN, CP 1641, 59072-970, Natal, RN, Brazil \\ 
$^{12}$ Department of Earth Sciences, University of Hawaii at Manoa, 1680 East-West Rd, Honolulu, HI 96822, USA 
}

\date{Accepted 2024 November 1. Received 2024 October 24; in original form 2024 August 24} 

\pubyear{2023}

\begin{document}

\label{firstpage}
\pagerange{\pageref{firstpage}--\pageref{lastpage}}
\maketitle

\begin{abstract}
This paper presents near-infrared spectropolarimetric and velocimetric observations of the young planet-hosting T~Tauri star PDS~70, collected with SPIRou at the 
3.6-m Canada-France-Hawaii Telescope from 2020 to 2024.  Clear Zeeman signatures from magnetic fields at the surface of PDS~70 are detected in our data set of 40 
circularly polarized spectra.  Longitudinal fields inferred from Zeeman signatures, ranging from $-116$ to 176~G, are modulated on a timescale of $3.008\pm0.006$~d, 
confirming that this is the rotation period of PDS~70.  Applying Zeeman-Doppler imaging to subsets of unpolarized and circularly polarised line profiles, we show 
that PDS~70 hosts low-contrast brightness spots and a large-scale magnetic field in its photosphere, featuring in particular a dipole component of strength 
200--420~G that evolves on a timescale of months.  From the broadening of spectral lines, we also infer that PDS~70 hosts a small-scale field of $2.51\pm0.12$~kG.  
Radial velocities derived from unpolarized line profiles are rotationally modulated as well, and exhibit additional longer-term chromatic variability, most likely 
attributable to magnetic activity rather than to a close-in giant planet (with a 3-$\sigma$ upper limit on its minimum mass of $\simeq$4~\mjup\ at a distance of 
$\simeq$0.2~au).  We finally confirm that accretion occurs at the surface of PDS~70, generating modulated red-shifted absorption in the 1083.3-nm \hei\ triplet, and 
show that the large-scale magnetic field, often strong enough to disrupt the inner accretion disc up to the corotation radius, weakens as the star gets fainter and 
redder (as in 2022), suggesting that dust from the disc more easily penetrates the stellar magnetosphere in such phases. 
\end{abstract}

\begin{keywords}
stars: magnetic fields --
stars: imaging --
stars: planetary systems --
stars: formation --
stars: individual:  PDS~70  --
techniques: polarimetric
\end{keywords}



\section{Introduction}
\label{sec:int}

Over the last few decades, ground-based and space-borne observations have revealed a wealth of information on how low-mass stars and their planets build up from 
giant molecular clouds under the combined effect of turbulence, gravitation and magnetic fields, first developing filamentary structures within which local 
overdensities, called dense cores, end up collapsing under their own weight \citep[e.g.,][]{Andre14}.  In this collapse, a rotating disc of gas and dust forms 
around the central mass, which progressively becomes a growing protostar through intense accretion from the disc, then a pre-main-sequence (PMS) star 
(also called T~Tauri star / TTS) at an age of a few Myr, once accretion weakens and the newborn star gets warm enough to clear out its surrounding dust envelope 
through radiation and outflows \citep[e.g.,][]{Tsukamoto23,Kuffmeier24,Cabrit24}.  Planets form more or less at the same time, with gas and dust particles within the disc 
colliding together and growing into pebbles, then planetesimals, before becoming actual protoplanets \citep[e.g.,][]{Drazkowska23}.  Giant planets presumably form in the 
outer, cooler regions of the protoplanetary disc, beyond the snow line, where they can accumulate large amounts of hydrogen and helium, and may start migrating inward within the disc 
before it dissipates at an age of 3--10~Myr, to eventually become close-in giant planets, shaping at the same time the architecture of planetary systems 
\citep[e.g.,][]{Mordasini12,Chambers18,Lau24}.  Magnetic fields play a key role at virtually all stages in this process, by affecting the dynamics and impacting fragmentation within the disc, by 
altering the way disc material is accreted onto the star and planets, and by influencing how planets form and migrate within the disc \citep[e.g.,][]{Pudritz19}.  

Whereas observations of TTSs with multiple techniques operating at various wavelengths have allowed us to investigate in detail the key physical mechanisms at work 
in star formation, including magnetospheric accretion and ejection phenomena taking place between the star and the accretion disc \citep{Hartmann16,Hussain18,Bouvier22}, 
much less is known from actual data about forming planets at an age of a few Myr, with only a few young planets detected so far around low-mass stars either directly 
through imaging \citep[e.g.,][]{Xie20,Langlois21}, or indirectly through transit photometry or velocimetry \citep[e.g.,][]{David19,Mann22}.  Whereas the first approach is 
technically tricky given the combination of high-angular resolution and high-contrast performance that it mandatorily requires, the second one is just as complex, 
the size of the photometric or velocimetric signals being vastly smaller, even for close-in giant planets, than the huge spectral variability that TTSs exhibit as 
a result of accretion and magnetic activity.  This is why only a handful of planets around stars younger than 20~Myr have reliably been identified by such techniques so far.  

Initially identified as a PMS star \citep{Gregorio92,Gregorio02} through its spectral features (in particular \hal\ emission and 670.7-nm Li absorption) and the 
infrared excess indicating the presence of an extended accretion disc, PDS~70 is a very interesting target in this respect, as outlined more extensively in Sec.~\ref{sec:par}.  
Featuring a highly structured accretion disc 
with a large central cavity hosting two massive protoplanets \citep{Keppler18, Haffert19}, PDS~70 is the only TTS so far with multiple directly-imaged forming 
planets and thereby a key object for studying the late phases of protoplanetary disc evolution and planet formation.  At an age of about $\simeq$5~Myr, both the star 
and the two distant giant planets {\emr (possibly even younger than the star)} are still in the process of accreting material from the surrounding disc, offering us a glimpse 
into stages of planetary development that are only observable in young systems and hard to infer from data on older systems.  PDS~70 thereby provides us with the opportunity of 
bridging the gap between theoretical models and observations of planet formation. 

Our paper is structured as follows.  After a short revisit of the main characteristics of PDS~70 in Sec.~\ref{sec:par}, we detail our new near-infrared (nIR) SPIRou 
observations in Sec.~\ref{sec:obs}, outline the spectropolarimetric results obtained for this star (in Sec.~\ref{sec:bl}) and the tomographic modeling inferred with 
Zeeman-Doppler Imaging (ZDI, see Sec.~\ref{sec:zdi}).  We also present the velocimetric study based on our SPIRou data (in Sec.~\ref{sec:rvs}) as well as a short analysis 
of the few nIR emission lines of PDS~70 (in Sec.~\ref{sec:eml}).  We conclude with a final summary of our results and a short discussion on what they tell us about 
star and planet formation (in Sec.~\ref{sec:dis}).

\section{The T~Tauri star PDS~70}
\label{sec:par}

PDS~70 (V1032~Cen, CD-40~8434, IRAS~14050-4109) is a TTS located at a distance of $112.4\pm0.3$~pc from the Sun \citep{Gaia23} in the Sco-Cen association.  
PDS~70 was first identified as a weak-line TTS \citep{Gregorio92,Gregorio02}, i.e., a TTS that experiences no accretion at the surface and thereby exhibits only weak emission lines.  
The highly-structured accretion disc of PDS~70 indeed features a large central dust-depleted cavity \citep{Hashimoto12}, within which two massive protoplanets were detected 
through direct imaging (at radii of 21 and 35~au), and shown to be accreting from a circumplanetary disc \citep{Keppler18,Haffert19,Thanathibodee19,Isella19,Benisty21,Blakely24}.  
PDS~70 itself was finally also proven to be accreting from an inner disc (of radius $\simeq$10~au), albeit at very low accretion rates \citep{Haffert19,Thanathibodee20}.   

The photospheric temperature of PDS~70 was initially measured at $\teff=4400\pm100$~K from fitting the spectral energy distribution \citep{Gregorio02}, then later revised 
to better match its $B-V=1.24$ color index, in agreement with a cooler temperature of $\teff=4020$~K \citep[with $\AV\simeq0$,][]{Pecaut13} or 4140~K (with $\AV\simeq0.2$).  
The putatively most accurate set of photospheric parameters were derived from the latest Gaia data (DR3), yielding $\teff=4140\pm10$~K and $\logg=4.15\pm0.01$ \citep{Gaia23}.  
Given the minimum $V$ magnitude (unspotted brightness) of PDS~70, equal to $V=12.05$ according to the ASAS-SN $V$-band photometry \citep[][see top panel of Fig.~\ref{fig:lc2} 
in Appendix~\ref{sec:appA}]{Kochanek17}, 
the quoted distance and photospheric temperature \citep[implying a bolometric correction of $-0.95$,][]{Pecaut13}, we infer for this star a logarithmic luminosity with respect 
to the Sun equal to $\log(\lstar/\lsun)=-0.36\pm0.04$ and thereby a radius of $\rstar=1.29\pm0.06$~\rsun.  
Besides, a precise mass estimate of $\mstar=0.875\pm0.030$~\msun\ was derived for PDS~70 from the disc gas kinematics observed with ALMA \citep{Keppler19}.  Coupled to 
the \logg\ estimate from Gaia, we can infer an independent radius estimate for PDS~70, equal to $\rstar=1.31\pm0.06$~\rsun\ and in good agreement with the previous one. 
In the following, we take the average between both estimates and thus assume that PDS~70 has a radius of $\rstar=1.30\pm0.06$~\rsun, implying at the same time 
$\log(\lstar/\lsun)=-0.35\pm0.04$.  

Comparing to the evolutionary models of \citet{Baraffe15}, we infer from \teff\ and \logg\ a stellar mass of $\mstar=0.86\pm0.01$~\msun, very close to the dynamical mass 
of \citet{Keppler19}, and an age of $5.8\pm0.3$~Myr, consistent with other literature estimates.  We note that in the particular case of PDS~70, the mass derived from 
the evolutionary models of \citet{Baraffe15} is not overestimated by $\simeq$30~per cent, as opposed to what was inferred for this mass range from several other PMS stars for 
which dynamical masses were also available \citep{Braun21}.  At this age, PDS~70 is most likely no longer fully convective, with about 20~per cent of the central mass forming a 
radiative core {\emr \citep[according to][]{Baraffe15}}.  

The rotation period we derive for PDS~70 from spectropolarimetry, equal to $\Prot=3.008\pm0.006$~d (see Sec.~\ref{sec:bl}), is consistent with the previously published 
estimate \citep{Thanathibodee20} and with that inferred from the ASAS-SN $V$-band and $g$-band light curves ($3.010\pm0.010$~d, see Sec.~\ref{sec:bl} and Fig.~\ref{fig:lc}).  
Once coupled to our measurement of the line-of-sight projected equatorial rotation velocity of PDS~70 $\vsini=19.0\pm0.5$~\kms\ derived from brightness and magnetic mapping 
with ZDI (see Sec.~\ref{sec:zdi}), it yields $\rstar \sin i = 1.13\pm0.03$~\rsun, thus implying an inclination of the rotation axis of PDS~70 to the line of sight of 
$i=60\degr\pm6\degr$.  This is slightly larger than (though still roughly consistent with) the inclination of the normal of the outer disc to the line of sight \citep[equal 
to $i_{\rm disc}=51.7\degr$,][]{Keppler19}, suggesting that any potential misalignment between the equatorial plane of PDS~70 and its surrounding disc should be small.   
The corotation radius, at which the Keplerian angular rotation rate equals the angular rotation rate at the surface of the star, is equal to 
$\rcor=0.039\pm0.001$~au ($6.5\pm0.3$~\rstar).  

The logarithmic mass accretion rate $\log\Mdot$ reported to take place at the surface of PDS~70 is low, ranging from $-10.3$ to $-9.7$ \citep[in \mspy, with an uncertainty of 
about 0.3,][]{Haffert19,Thanathibodee20,Skinner22,Campbell23,Gaidos24}.  
Although weak, accretion can 
significantly distort the light curve, with rotational modulation getting weaker (see, e.g., the stacked periodogram of the ASAS-SN $V$-band and $g$-band light curves in 
Fig.~\ref{fig:lc}), sometimes to the point of vanishing, as in the sector 38 and 65 TESS light curves \citep{Gaidos24}.  Using the wind-tracing 630-nm \foxi\ line, a 
magnetically-driven disc wind was also detected in the inner disc of PDS~70, originating from a radius of 0.1--0.2~au \citep{Campbell23}.  
We come back on these points in Sec.~\ref{sec:eml}.  

All stellar parameters used in or derived from this study are listed in Table~\ref{tab:par}.  

\begin{table}
\caption[]{Parameters of PDS~70 used in or derived from our study} 
\scalebox{0.95}{\hspace{-4mm}
\begin{tabular}{ccc}
\hline
distance (pc)        & $112.4\pm0.3$   & \citet{Gaia23} \\
\teff\ (K)           & $4140\pm10$     & \citet{Gaia23} \\
\logg\ (dex)         & $4.15\pm0.01$   & \citet{Gaia23} \\ 
\mstar\ (\msun)      & $0.875\pm0.030$ & \citet{Keppler19} \\
\rstar\ (\rsun)      & $1.30\pm0.06$   & \\
$\log(\lstar/\lsun)$ & $-0.35\pm0.04$  & \\ 
age (Myr)            & $5.8\pm0.3$     & \citet{Baraffe15} \\ 
\Prot\ (d)           & $3.010$         & period used to phase data \\ 
\Prot\ (d)           & $3.008\pm0.006$ & from \Bl\ data \\ 
\Prot\ (d)           & $3.010\pm0.009$ & from ASAS-SN $V$ and $g$ data \\ 
\vsini\ (\kms)       & $19.0\pm0.5$    & from ZDI modeling \\ 
$\rstar \sin i$ (\rsun)& $1.13\pm0.03$ & from \rstar\ and \vsini \\ 
$i$ (\degr)          & $60\pm6$        & from \rstar\ and $\rstar \sin i$ \\ 
$i_{\rm disc}$ (\degr) & $51.7\pm0.1$  & \citet{Keppler19} \\ 
<$B$> (kG)           & $2.51\pm0.12$   & on median spectrum \\
\rcor\ (au)          & $0.039\pm0.001$ & from \mstar\ and \Prot \\ 
\rcor\ (\rstar)      & $6.5\pm0.3$     & from \rcor\ (au) and \rstar \\ 
$\log\Mdot$ (\mspy)  & $-10.3$ to $-9.7$ & \citet{Thanathibodee20} \\ 
\hline
\end{tabular}}
\label{tab:par}
\end{table}

\section{SPIRou observations}
\label{sec:obs}

We observed PDS~70 in 3 different seasons with the SPIRou nIR spectropolarimeter / high-precision velocimeter \citep{Donati20} at CFHT, within the 
SPIRou Legacy Survey (SLS) in 2020 May-June, and 2022 June, then within the SPICE Large Programme in 2024 March-May.  SPIRou collects unpolarized and polarized stellar 
spectra, covering a wavelength interval of 0.95--2.50~\mic\ at a resolving power of 70\,000 in a single exposure.  In this study, we focussed on 
circularly polarized (Stokes $V$) and unpolarized (Stokes $I$) spectra only.  Polarization observations usually consist of sequences of 4 
sub-exposures, with each sub-exposure corresponding to a different azimuth of the Fresnel rhomb retarders of the SPIRou polarimetric unit.  This 
procedure was shown to succeed in removing systematics in polarization spectra \citep[to first order, see, e.g.,][]{Donati97b}.  Each recorded 
sequence yields one Stokes $I$ and one Stokes $V$ spectrum, as well as one null polarization check (called $N$) used to diagnose potential 
instrumental or data reduction issues.  In case of unstable weather conditions where only 2 sub-exposures can be recorded, one can still retrieve 
a pair of Stokes $I$ and $V$ spectra, but no $N$ spectrum and a slightly worse compensation of systematics.  

We recorded a total of 43 polarization sequences for PDS~70, 4 in 2020, 10 in 2022 and 29 in 2024, with a single 
sequence collected in most clear nights.  As a result of poor weather leading to very low signal to noise ratios (SNRs), 3 spectra were discarded, 
one in the second season (on June 13) and two in the third one (on April 30 and May 28);  weather issues also caused one sequence of the second season 
(on June 12) to include only 2 subexposures.  Our final data set thus included 40 validated Stokes $I$ and $V$ spectra of PDS~70 over 3 seasons, with 
seasonal subsets respectively spanning intervals of 27~d (4 points), 10~d (9 points) and 71~d (27 points), and altogether covering a temporal window 
of 1480~d.  The full log of our observations is provided in Table~\ref{tab:log} in Appendix~\ref{sec:appB}.   

All SPIRou spectra of PDS~70 were processed with \texttt{Libre ESpRIT}, the nominal reduction pipeline of ESPaDOnS at CFHT, optimized for 
spectropolarimetry and adapted for SPIRou \citep{Donati20}.  Subsequently, we applied Least-Squares Deconvolution \citep[LSD,][]{Donati97b} to the 
reduced spectra, with a line mask computed with the VALD-3 database \citep{Ryabchikova15} for a set of atmospheric parameters ($\teff=4000$~K and 
$\logg=4.0$) matching those of PDS~70 (see Sec.~\ref{sec:par}).  As in previous studies, we only selected atomic lines deeper than 10 percent of the 
continuum level, for a total of $\simeq$1500 lines of average wavelength and Land\'e factor equal to 1750~nm and 1.2 respectively.  The noise levels 
$\sigma_V$ in the resulting Stokes $V$ LSD profiles range from 1.9 to 4.8 (median 2.2, in units of $10^{-4} I_c$ where $I_c$ denotes the continuum 
intensity).  These Stokes $I$ and $V$ LSD profiles were mostly used to investigate the magnetic properties of PDS~70 (see Secs.~\ref{sec:bl} and 
\ref{sec:zdi}) as well as its velocimetric behaviour (see Sec.~\ref{sec:rvs}), and to estimate the spectral veiling (in the $J$ and $H$ bands mostly) 
reducing the depth of spectral lines as a result of accretion (see Sec.~\ref{sec:eml}).  We also constructed a second LSD mask, containing the 
$\simeq$300 CO lines of the CO bandhead only, known to be insensitive to magnetic fields, to help diagnose the impact of activity on radial velocities 
(RVs, see Sec.~\ref{sec:rvs}) and measure the veiling in the $K$ band (see Sec.~\ref{sec:eml}). 

Phases and rotation cycles are derived assuming a rotation period of $\Prot=3.010$~d (see Table~\ref{tab:par}) and counting from an 
arbitrary starting BJD0 of 2458977.9 (i.e., prior to our first SPIRou observation).

\section{Spectropolarimetric analysis}
\label{sec:bl}

The inferred Stokes $I$ and $V$ LSD profiles of PDS~70 show clear Zeeman signatures within spectral lines.  This is best evidenced by deriving the 
longitudinal field \Bl, i.e., the line-of-sight projected component of the vector field at the surface of the star averaged over the visible stellar 
hemisphere, associated with each pair of Stokes $V$ and $I$ LSD profiles following \citet{Donati97b}.  In practice, we achieve this by computing the 
first moment of the Stokes $V$ profile and its error bar, which we normalise with the equivalent width of the Stokes $I$ LSD profiles estimated 
through a standard Gaussian fit.  Given the \vsini\ of PDS~70 (of $19.0\pm0.5$~\kms, see Sec.~\ref{sec:zdi}), the first moment of Stokes $V$ LSD profiles 
were computed on a window of $\pm34$~\kms\ in the stellar rest frame in order not to miss some of the polarimetric information and yet minimize the impact 
of photon noise.  We find that \Bl\ varies from $-116$ to 176~G (median 66~G) over our observing campaign with error bars ranging from 10 to 30~G 
(median 13~G), implying a reduced chi-square \chisqr\ (with respect to the $\Bl=0$~G level) equal to $\chisqr=47.9$ and thus a very clear magnetic detection.  
Processing the polarization check $N$ in the same way yields $\chisqr=1.16$, consistent with no signal and indicating that our spectropolarimetric 
measurements are free of spurious pollution and that our formal error bars agree with the observed measurement dispersion.  

\begin{figure*}
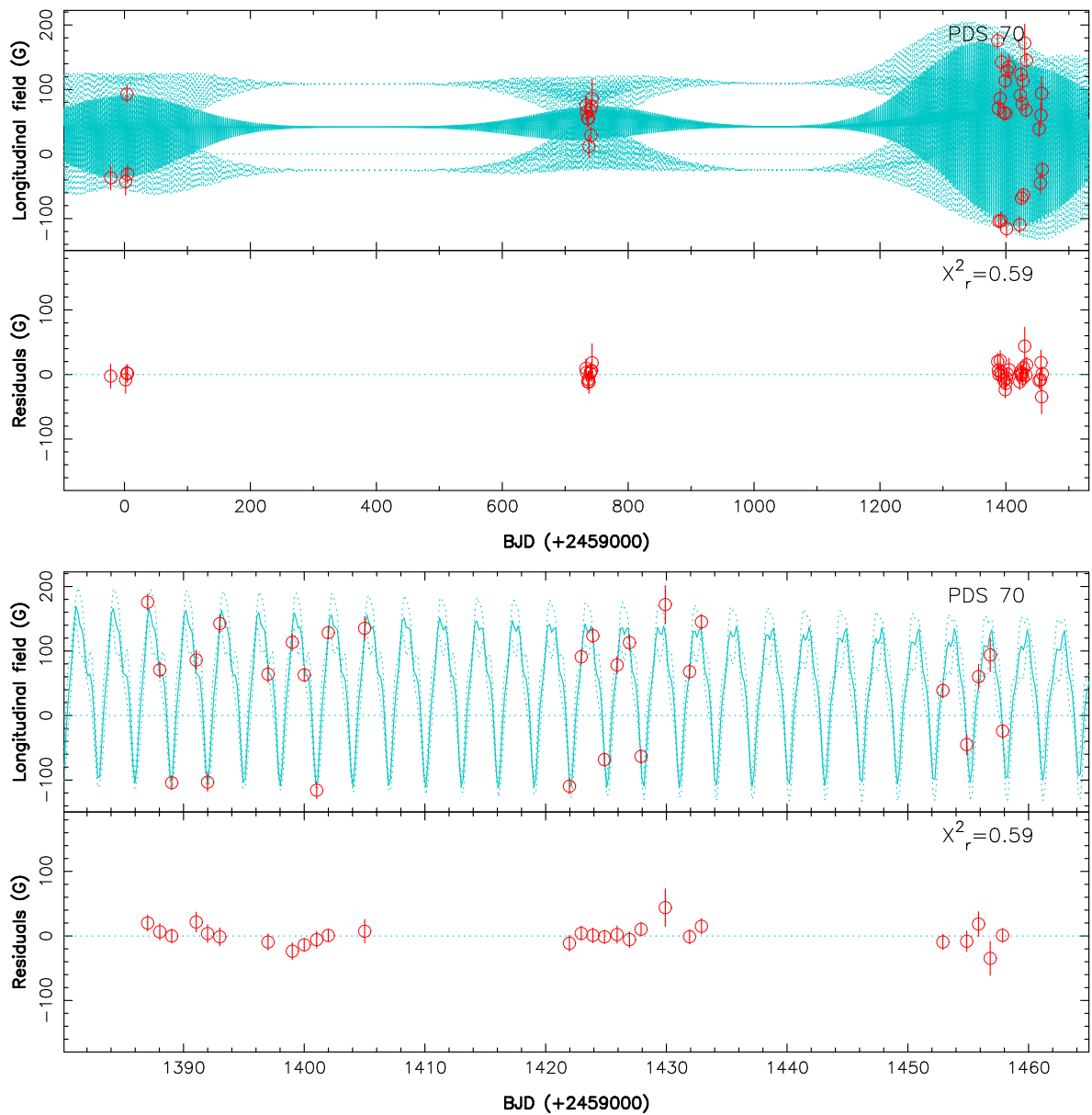

\centerline{\includegraphics[scale=0.6,angle=-90]{fig/pds70-bl1.ps}\vspace{2mm}}
\centerline{\includegraphics[scale=0.6,angle=-90]{fig/pds70-bl2.ps}}
\caption[]{Longitudinal magnetic field \Bl\ of PDS~70 (red open circles) as measured with SPIRou, and QP GPR fit to the data (cyan full line) with 
corresponding 68~per cent confidence intervals (cyan dotted lines).  The top panel shows the full data set whereas the bottom panel is a zoom on the 2024 
(March to May) observations.  The residuals, featuring an rms of 10.1~G ($\chisqr=0.59$), are shown in the bottom plot of each panel.  The 3-d rotational modulation 
is clear in all seasons, especially in the last one where the amplitude of the modulation is largest.  } 
\label{fig:bl}
\end{figure*}

To investigate rotational modulation of our \Bl\ data, allowing for temporal evolution of the modulation pattern, we proceeded as in our previous 
studies, i.e., by using quasi-periodic (QP) Gaussian-Process Regression (GPR).  This is achieved by finding out the hyper parameters of the covariance 
function that best describes our \Bl\ data, arranged in a vector denoted $\bf y$.  The QP covariance function $c(t,t')$ we use in this purpose is as 
follows: 
\begin{eqnarray}
c(t,t') = \theta_1^2 \exp \left( -\frac{(t-t')^2}{2 \theta_3^2} -\frac{\sin^2 \left( \frac{\pi (t-t')}{\theta_2} \right)}{2 \theta_4^2} \right) 
\label{eq:covar}
\end{eqnarray}
where $\theta_1$ is the amplitude (in G) of the Gaussian Process (GP), $\theta_2$ its recurrence period (directly linked to \Prot), $\theta_3$ the 
evolution timescale (in d) on which the shape of the \Bl\ modulation changes, and $\theta_4$ a smoothing parameter describing the amount of harmonic 
complexity needed to describe the data \citep{Haywood14,Rajpaul15}  
We then select the set of hyper parameters that yields the highest likelihood $\mathcal{L}$, defined by: 
\begin{eqnarray}
2 \log \mathcal{L} = -n \log(2\pi) - \log|{\bf C+\Sigma+S}| - {\bf y^T} ({\bf C+\Sigma+S})^{-1} {\bf y}
\label{eq:llik}
\end{eqnarray}
where $\bf C$ is the covariance matrix for our 40 epochs, $\bf \Sigma$ the diagonal variance matrix associated with $\bf y$, and 
${\bf S}=\theta_5^2 {\bf J}$ ($\bf J$ being the identity matrix) the contribution from an additional white noise source used as a fifth 
hyper-parameter $\theta_5$ (in case \Bl\ is affected by intrinsic variability beyond the photon noise described by our formal error bars).  
We then use a Monte-Carlo Markov Chain (MCMC) process to explore the hyper parameter domain, yielding posterior distributions and error bars for 
each of them.  We use here our MCMC and GPR modeling tools, as in our previous studies \citep[e.g.,][]{Donati23,Donati23b,Donati24,Donati24b}.  
{\emr The MCMC process is a conventional single chain Metropolis-Hastings scheme, typically carried out over a few $10^5$ steps, including the first 
few $10^4$ steps as burn-in, with convergence estimated from an autocorrelation analysis. } 
The marginal logarithmic likelihood $\log \mathcal{L}_M$ of a given solution is computed using the approach of \citet{Chib01} as in, e.g., 
\citet[][]{Haywood14}.  

We started the exploration with all 5 hyper parameters free to vary, then ended up fixing $\theta_3$ and $\theta_4$, {\emr weakly constrained by our small 
set of observations, to the values (100~d and 0.3, see Table~\ref{tab:bl}) yielding a marginal improvement in $\log \mathcal{L}$, with virtually no impact 
on the results and in particular on the derived \Prot.  This is due to both the small number of data points in any given season (limiting 
the achievable precision on $\theta_3$) and the rather sparse sampling of the rotation cycle (\Prot\ being very close to a small integer number of days, 
limiting the precision on $\theta_4$)}.  
The fit we obtain, shown in Fig.~\ref{fig:bl} and matching our data at a level of $\chisqr=0.59$ {\emr (taking into account photon noise only and not $\theta_5$)},  
yields the hyper parameters listed in Table~\ref{tab:bl} and an estimate of $\Prot=3.008\pm0.006$~d.  
This value is consistent within error bars with the one derived from the same treatment of the ASAS-SN $V$-band and $g$-band light curves (with all 5 hyper 
parameters free to vary), giving $\Prot=3.010\pm0.009$~d.   The achieved precision on the derived rotation period only allows phase consistency to be 
ensured within a single season, as a result of both the temporal distribution of our data and the limited lifetime of the magnetic features at the 
surface of PDS~70 (see below).  

We note that the full amplitude of the modulation of \Bl\ {\emr varied from season to season, being larger in 2024 (290~G) than in 
the 2 previous seasons and in 2022 in particular (70~G) where the Stokes $V$ signatures were weak over a large fraction of the rotation cycle (see  
Sec.~\ref{sec:zdi}).  The \Bl\ curve also changed in both shape and amplitude over the 3 SPIRou runs of 2024 (see bottom panel of Fig.~\ref{fig:bl}).  
This seasonal magnetic evolution, only marginally significant when considering \Bl\ alone ($\theta_3$ being only weakly constrained by our observations), 
is nonetheless clear from the temporal evolution of the LSD Stokes $V$ profiles, as emphasized in Sec.~\ref{sec:zdi}.} 
It tells us that the large-scale field of PDS~70 evolves with time as expected from a dynamo field, and does it on a timescale of order of a few months, 
similar to (though not as fast as) what was reported for the other planet-hosting non-fully-convective PMS star V1298~Tau \citep{Finociety23b}.  

\begin{table} 
\caption[]{Results of our MCMC modeling of the \Bl\ curve of PDS~70 with QP GPR.  For each fitted hyper-parameter, we list the derived value, the 
corresponding error bar and the assumed prior.  {\emr Using a uniform prior (over an interval of 2.5--3.5~d) for the recurrence period virtually yields  
the same results.}  The knee of the modified Jeffreys prior is set to the median error bar of our 
\Bl\ measurements $\sigma_{B}$, equal to 13~G.  We also quote the resulting \chisqr\ and rms of the final GPR fit. }  
\scalebox{0.95}{\hspace{-6mm}
\begin{tabular}{cccc}
\hline
Parameter   & Name & Value & Prior   \\
\hline 
Amplitude (G)        & $\theta_1$  & $70^{+15}_{-13}$  & mod Jeffreys ($\sigma_{B}$) \\
Rec.\ period (d)     & $\theta_2$  & $3.008\pm0.006$   & Gaussian (3.0, 0.2) \\
Evol.\ timescale (d) & $\theta_3$  & 100               & fixed \\
Smoothing            & $\theta_4$  & 0.30              & fixed \\
White noise (G)      & $\theta_5$  & $5^{+5}_{-3}$     & mod Jeffreys ($\sigma_{B}$) \\
\hline
\chisqr              &  & 0.59 &  {\emr including photon noise only}   \\ 
rms (G)              &  & 10.1 &              \\ 
\hline 
\end{tabular}}      
\label{tab:bl}      
\end{table}          

\begin{figure}
\centerline{\includegraphics[scale=0.6,bb=20 20 380 390]{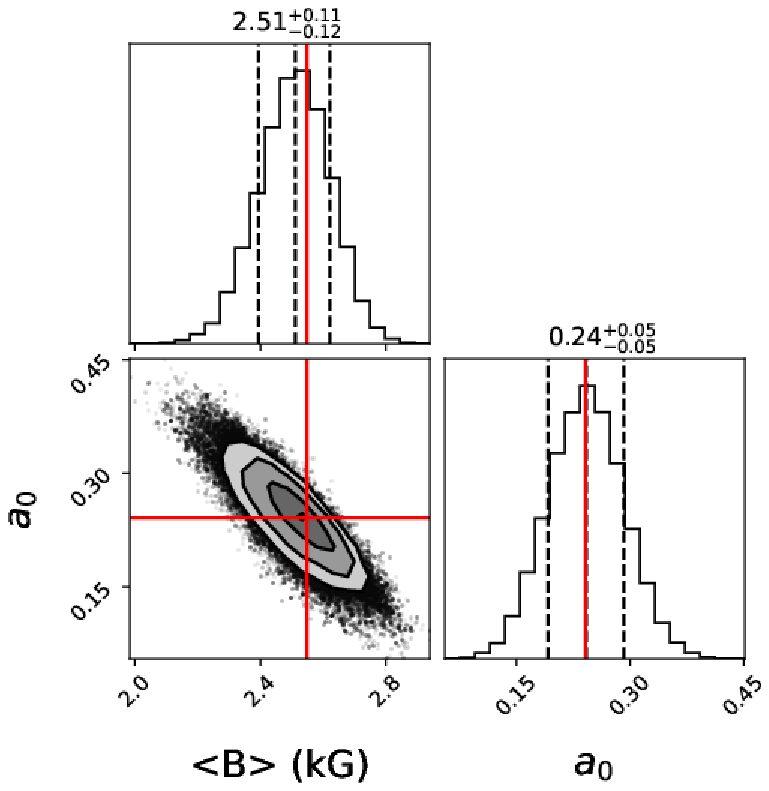}}
\caption[]{Magnetic parameters of PDS~70, derived by fitting our median SPIRou spectrum using the modeling approach of \citet{Cristofari23},
which incorporates magnetic fields as well as a MCMC process to determine optimal parameters and their error bars.  We find that PDS~70 hosts a 
small-scale magnetic field of <$B$>~$=2.51\pm0.12$~kG whereas the relative area of non-magnetic regions is $a_0=24\pm5$~per cent, {\emr demonstrating 
a clear detection}.  }
\label{fig:bmag}
\end{figure}

We finally looked at the Zeeman broadening of atomic and molecular lines with ZeeTurbo \citep{Cristofari23,Cristofari23b} applied to our median spectrum 
of PDS~70, and find that 4 magnetic components, associated with small-scale magnetic fields of strengths 0, 2, 4 and 6~kG, and respective filling factors 
$a_0=24\pm5$~per cent, $a_2=45\pm7$~per cent, $a_4=11\pm7$~per cent and $a_6=19\pm4$~per cent of the visible stellar surface, are needed to obtain a good 
match to the spectrum, yielding an overall small-scale field measurement of <$B$>~$=2.51\pm0.12$~kG for PDS~70 ({\emr see the derived corner plot of <$B$> 
vs $a_0$ in Fig.~\ref{fig:bmag}}).  {\emr Adding more parameters in the modeling (i.e., for a 8~kG or even a 10~kG component) only marginally changes <B> 
(by about 0.1~kG).} 

\section{Zeeman-Doppler Imaging}
\label{sec:zdi}

We analysed the Stokes $I$ and $V$ LSD profiles of PDS~70 using ZDI to simultaneously reconstruct the topology of the 
large-scale magnetic field and the distribution of brightness features at the surface of the star.  Given the relatively fast temporal evolution of \Bl\ 
(see Sec.~\ref{sec:bl}), we split our data into 5 subsets, respectively corresponding to epochs 2020 May-June, 2022 June, 2024 March, 2024 April and 2024 
May, each gathering data collected over a few (2--9) rotation cycles.  We stress again that the period of PDS~70, very close to a small integer number of 
days, does not allow one to easily obtain a dense sampling of the rotation cycle.  {\emr This limitation is partly compensated by the relatively low limb 
darkening \citep[$\simeq$0.4 in the H band,][]{Claret95} and the significant \vsini, with each Doppler-broadened LSD profile yielding information about 9 
independent velocity strips across the stellar disc.  As a result, we estimate that the spatial localisation of brightness features is still reasonably accurate
(typically 10\degr\ in longitude and 20\degr\ in latitude), with most reconstructed features detected either as they cross the meridian or between 
two neighbouring phases of the rotation cycle. } 
However, with only 4 spectra spread on about 40~per cent of the rotation cycle, the first of our 5 subsets is clearly too sparse to yield fully reliable ZDI maps.  
We nonetheless derived our results in the same way for all epochs, and discuss limitations whenever relevant.  

For this study, we used the same ZDI code as in previous similar analyses of SPIRou data \citep[e.g.,][]{Donati23,Finociety23b}, which allows one 
to reconstruct the brightness distribution and the topology of the large-scale magnetic field at the surface of a rotating star from phase-resolved 
sets of Stokes $I$ and $V$ LSD profiles.  This is performed through an iterative process, starting from a small magnetic field and a featureless 
brightness map and progressively adding information at the surface of the star, exploring the parameter space with a kind of conjugate gradient 
technique, until the modeled Stokes $I$ and $V$ profiles match the observed ones at the required level, usually $\chisqr\simeq1$ \citep[see, 
e.g.,][for more information on ZDI]{Brown91, Donati97c, Donati06b}.  As this inversion problem is ill-posed, ensuring a unique solution mandatorily 
requires regularization, which in our case follows the principles of maximum entropy image reconstruction \citep{Skilling84} to select the image 
featuring minimal information among those matching the data.  

\begin{figure*}
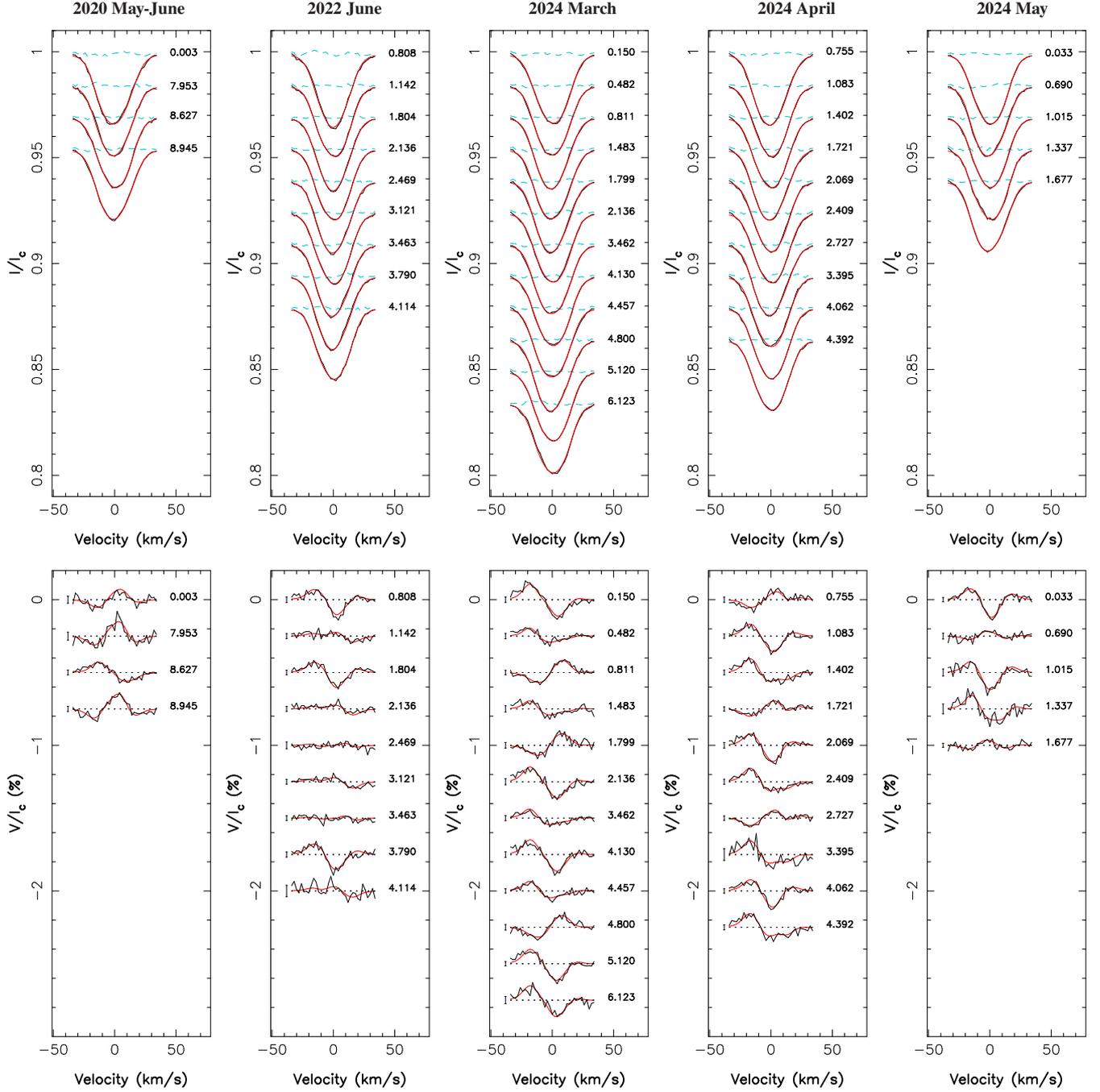

\flushleft{\bf \hspace{1.1cm}2020 May-June\hspace{2cm}2022 June\hspace{2.4cm}2024 March\hspace{2.3cm}2024 April\hspace{2.3cm}2024 May\vspace{-3mm}} 
\flushleft{\includegraphics[scale=0.49,angle=-90]{fig/pds70-i20-2.ps}\hspace{2mm}\includegraphics[scale=0.49,angle=-90]{fig/pds70-i22-2.ps}\hspace{2mm}\includegraphics[scale=0.49,angle=-90]{fig/pds70-i24a-2.ps}\hspace{2mm}\includegraphics[scale=0.49,angle=-90]{fig/pds70-i24b-2.ps}\hspace{2mm}\includegraphics[scale=0.49,angle=-90]{fig/pds70-i24c-2.ps}\vspace{1mm}} 
\flushleft{\includegraphics[scale=0.49,angle=-90]{fig/pds70-v20.ps}\hspace{2mm}\includegraphics[scale=0.49,angle=-90]{fig/pds70-v22.ps}\hspace{2mm}\includegraphics[scale=0.49,angle=-90]{fig/pds70-v24a.ps}\hspace{2mm}\includegraphics[scale=0.49,angle=-90]{fig/pds70-v24b.ps}\hspace{2mm}\includegraphics[scale=0.49,angle=-90]{fig/pds70-v24c.ps}} 
\caption[]{Observed (thick black line) and modelled (thin red line) LSD Stokes $I$ (top row) and $V$ (bottom row) profiles of PDS~70 for epochs 2020 May-June, 2022 June, 
2024 March, 2024 April and 2024 May (left to right).  Observed profiles were derived by applying LSD to our SPIRou spectra, using the atomic line mask outlined in 
Sec.~\ref{sec:obs}.  Rotation cycles (counting from 0, 250, 468, 479 and 490 for the 5 epochs respectively, see Table~\ref{tab:log}) are indicated to the right of LSD 
profiles, while $\pm$1$\sigma$ error bars are added to the left of Stokes $V$ signatures.  {\emr In the top row, we also show the differences between the observed 
and modelled LSD Stokes $I$ profiles as cyan dashed lines.}  }
\label{fig:fit}
\end{figure*}

In practice, the surface of the star is described as a grid of 5000 cells, the spectral contributions of which are computed using Unno-Rachkovsky's
analytical solution of the polarized radiative transfer equation in a plane-parallel Milne Eddington atmosphere \citep{Landi04}, with a local profile 
centred on 1750~nm and featuring a Doppler width and Land\'e factor of 3~\kms\ and 1.2 respectively \citep[as in our previous studies, e.g.,][]{Donati24b}.  
By summing up the contributions of all grid cells, taking into account the star and cell characteristics and assuming the star rotates as a solid body, 
one can compute the synthetic profiles at the observed rotation cycles.  
Whereas the relative brightness at the surface of the star is simply described as a series of independent pixels, the large-scale magnetic field is 
expressed as a spherical harmonics expansion, using the formalism of \citet[][see also \citealt{Lehmann22,Finociety22,Donati23}]{Donati06b} in which 
the poloidal and toroidal components of the vector field depend on 3 sets of complex coefficients, $\alpha_{\ell,m}$ and $\beta_{\ell,m}$ for the 
poloidal component, and $\gamma_{\ell,m}$ for the toroidal component, where $\ell$ and $m$ note the degree and order of the corresponding spherical 
harmonic term in the expansion.  Given the significant \vsini\ of PDS~70, we used a spherical harmonic expansion with terms up to $\ell=10$ in our study. 

\begin{figure*}
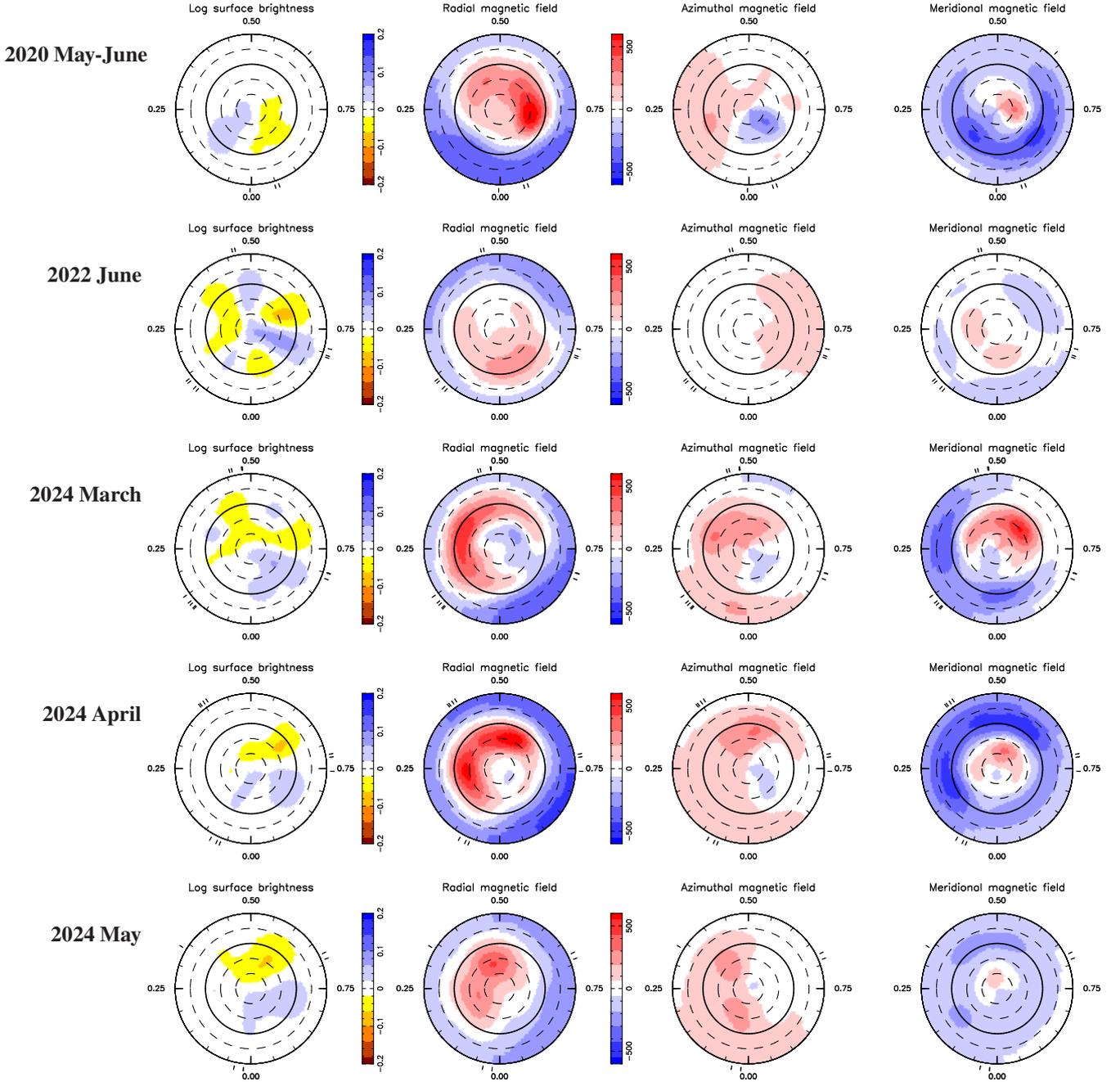

\flushright{\large\bf 2020 May-June\raisebox{0.3\totalheight}{\includegraphics[scale=0.42,angle=-90]{fig/pds70-map20.ps}}\vspace{1mm}}
\flushright{\large\bf 2022 June    \raisebox{0.3\totalheight}{\includegraphics[scale=0.42,angle=-90]{fig/pds70-map22.ps}}\vspace{1mm}}
\flushright{\large\bf 2024 March  \raisebox{0.3\totalheight}{\includegraphics[scale=0.42,angle=-90]{fig/pds70-map24a.ps}}\vspace{1mm}}
\flushright{\large\bf 2024 April  \raisebox{0.3\totalheight}{\includegraphics[scale=0.42,angle=-90]{fig/pds70-map24b.ps}}\vspace{1mm}}
\flushright{\large\bf 2024 May    \raisebox{0.3\totalheight}{\includegraphics[scale=0.42,angle=-90]{fig/pds70-map24c.ps}}}
\caption[]{Reconstructed maps of the brightness and large-scale magnetic field of PDS~70 for epochs 2020 May-June, 2022 June, 2024 March, 2024 April and 2024 May 
(top to bottom), derived from the Stokes $I$ and $V$ LSD profiles of Fig.~\ref{fig:fit} using ZDI.  In each row and from left to right, we show the logarithmic 
relative brightness with respect to the quiet photosphere, and the radial, azimuthal and meridional field components in spherical coordinates (in G).  
All maps are shown in a flattened polar projection down to latitude $-60$\degr, with the north pole at the center and the equator depicted as a bold line.  
Outer ticks indicate phases of observations.  Bright and dark spots respectively appear as blue and yellow/orange features, whereas positive radial, azimuthal and 
meridional fields respectively point outwards, counterclockwise and polewards.  } 
\label{fig:map}
\end{figure*}

Before applying ZDI, all Stokes $I$ LSD profiles of PDS~70 were corrected for the moderate veiling (see Sec.~\ref{sec:eml}) by normalizing them to the same
equivalent width (EW), while Stokes $V$ profiles were scaled accordingly.  
Running ZDI on our 5 subsets, we obtained the fits shown in Fig.~\ref{fig:fit} and the brightness and magnetic maps depicted in Fig.~\ref{fig:map}.  As 
a by product, we derived a measurement of \vsini, equal to $\vsini=19.0\pm0.5$~\kms, slightly larger than the estimate of \citet{Thanathibodee20} derived 
from much lower SNR HARPS spectra.  

The recovered brightness maps only show low contrast features, as often the case for cool stars in the nIR \citep{Finociety22,Finociety23b,Donati24} 
and in agreement with the Stokes $I$ LSD profiles of PDS~70 exhibiting low level distortions only (see Fig.~\ref{fig:fit} top panel).  The reconstructed features are 
nonetheless most likely real, given how they affect RVs (see Sec.~\ref{sec:rvs}) and how they repeatedly appear in the 3 independent images derived from the 2024 
subsets.  The low-latitude brightness inhomogeneities visible in these 3 maps around phase 0.75 (see Fig.~\ref{fig:map}) even suggest that differential rotation 
is likely present at the surface of PDS~70, progressively shifting these features to lower phases as time passes.  The corresponding phase shift (0.1~cycle in a 
timescale of 60~d) yields a differential rotation rate of order 10~\mrpd\ at lower latitudes with respect to the average rotation rate used to phase our 
data\footnote{A more quantitative analysis, with differential rotation used as a free ZDI parameter \citep[as in, e.g.,][]{Donati03b}, is not possible in the 
present case given {\emr the sparse sampling of the rotation cycle} and the strong intrinsic variability that the large-scale field of PDS~70 is subject to.}.  
We come back on this point in Secs.~\ref{sec:rvs}, \ref{sec:eml} and \ref{sec:dis}.  

Regarding the magnetic topology, the ZDI results confirm our preliminary impression from the \Bl\ data that the large-scale field of PDS~70 evolves on a timescale 
of a few months. {\emr We are indeed unable to fit with ZDI our complete data set over all seasons, nor a subset covering two consecutive seasons, nor even that 
including all our 2024 data, down to the same level of precision, unambiguously demonstrating that the magnetic topology of PDS~70 significantly evolved with time throughout our 
observing campaign.}  More specifically, the average strength of the field we recover varies from epoch to epoch by a factor of about $2.5\times$, from 
160 to 395~G (see Table~\ref{tab:mag}), with the weakest reconstructed field in 2022 June (when Stokes $V$ signatures have low amplitudes for most of the rotation 
cycle, see Fig.~\ref{fig:fit}) and the strongest in 2024 April.  Note that given the limited phase coverage of our first subset, the corresponding large-scale field 
we retrieved, already the second strongest our our sample, is likely underestimated.  The poloidal component of 
the field is dominant, always larger than 70~per cent and most of the time of order 90~per cent of the reconstructed magnetic energy.  The dipole component 
of the poloidal field is the strongest, totalling between 46~per cent and 89~per cent of the poloidal field energy, and ranging between 200~G (in 2022) and about 
400~G (in 2020 and 2024), with an average of 320~G over our observing epochs.  As already pointed out from the \Bl\ curve (see Sec.~\ref{sec:bl}), the evolution of the 
large-scale field in 2024 is significant, with the dipole inclination to the rotation axis varying by about 25\degr\ between March and April, and its 
strength increasing from 280 to 420~G before decreasing again to 300~G in May.  Although some of this variation may reflect the different number of spectra 
in each subset and the limited sampling, we believe that most of it is real as the phase coverage is similar for all subsets (except the first one from 2020) given the value of \Prot.  

\begin{table} 
\caption[]{Properties of the large-scale magnetic field of PDS~70 for our 5 data subsets.  
We list the average reconstructed large-scale field strength <$B_V$> (column 2), the polar strength of the dipole 
component \Bd\ (column 3), the tilt of the dipole field to the rotation axis and the phase towards which it is tilted (column 4), and the amount of magnetic energy reconstructed in the poloidal component of 
the field and in the axisymmetric modes of this component (column 5).  Error bars (estimated from variational analyses) are typically equal to 10~per cent for field strengths and percentages, 10\degr\ for 
field inclinations, and about twice worse for the first subset (for which ZDI is less reliable given the sparse phase coverage). } 
\begin{tabular}{ccccc}
\hline
Data subset   & <$B_V$> & \Bd     & tilt / phase    & poloidal / axisym \\ 
              &  (G)    &  (G)    & (\degr)         & (per cent)              \\ 
\hline
2020 May-June & 345 & 395 & 37 / 0.59 & 93 / 67 \\  
2022 June     & 160 & 200 & 37 / 0.99 & 88 / 61 \\  
2024 March    & 320 & 280 & 57 / 0.32 & 71 / 57 \\  
2024 April    & 395 & 420 & 31 / 0.35 & 92 / 81 \\  
2024 May      & 245 & 300 & 33 / 0.34 & 86 / 70 \\  
\hline 
\end{tabular}
\label{tab:mag}
\end{table}

\section{RV analysis}
\label{sec:rvs}

We derived RVs of PDS~70 from our SPIRou Stokes $I$ profiles, to investigate in particular whether there may be close-in massive planets orbiting within a few 0.1~au of the 
host star,  in addition to the two distant giant planets discovered by direct imaging \citep{Keppler18, Haffert19}.  To measure RVs from Stokes $I$ LSD profiles, we proceeded 
as in our study of V347~Aur \citep{Donati24c}, i.e., by describing each individual LSD profile as a simple first order Taylor 
expansion constructed from the median of all LSD profiles.   This approach is similar to that used in the ``line-by-line'' technique \citep{Artigau22}, applied here to 
LSD profiles rather than to individual lines {\emr (not adequate for stars with large \vsini's experiencing accretion and veiling, 
see Sec.~\ref{sec:eml})}, and gives results that are slightly more precise and stable than a straightforward Gaussian fit to the LSD profiles.   
The derived RVs, ranging from 5.3 to 7.3~\kms\ with a median error bar of 0.16~\kms\ (see Table~\ref{tab:log}), exhibit an rms of 0.49~\kms, about 3$\times$ larger 
than the median error bar.  

\begin{table*} 
\caption[]{MCMC results for the two cases (activity only, activity plus putative planet) of our RV analysis of PDS~70, for atomic lines (columns 2 and 3) and for 
CO bandhead lines (columns 4 and 5). In each case, we list the recovered GP and putative planet parameters with their error bars, as well as the priors used 
whenever relevant.  The last 4 rows give the \chisqr\ and the rms of the best fit to our RV data, as well as the associated marginal logarithmic likelihood, 
$\log \mathcal{L}_M$, and marginal logarithmic likelihood variation, $\Delta \log \mathcal{L}_M = \log {\rm BF}$, with respect to the model without putative planet.  
Two GPR hyper-parameters, weakly constrained by our sparse data, were fixed to a typical value yielding optimal results in a preliminary run with all GPR variables 
free to vary, with virtually no difference on the outcome.  }
\begin{tabular}{cccccc}
\hline
                   & \multicolumn{2}{c}{Atomic lines}                & \multicolumn{2}{c}{CO bandhead}                 &         \\
Parameter          & activity only          & activity + planet      & activity only          & activity + planet      &   Prior \\
\hline
$\theta_1$ (\kms)  & $0.41^{+0.12}_{-0.09}$ & $0.44^{+0.14}_{-0.10}$ & $0.26^{+0.10}_{-0.07}$ & $0.21^{+0.08}_{-0.06}$ & mod Jeffreys ($\sigma_{\rm RV}$) \\
$\theta_2$ (d)     & $3.00\pm0.01$          & $2.99\pm0.01$          & $3.00\pm0.02$          & $3.00\pm0.02$          & Gaussian (3.0, 0.2) \\
$\theta_3$ (d)     & 150                    & 150                    & 150                    & 150                    &  \\
$\theta_4$         & 0.3                    & 0.3                    & 0.3                    & 0.3                    &  \\
$\theta_5$ (\kms)  & $0.22^{+0.05}_{-0.04}$ & $0.16^{+0.06}_{-0.04}$ & $0.12^{+0.08}_{-0.05}$ & $0.08^{+0.06}_{-0.04}$ & mod Jeffreys ($\sigma_{\rm RV}$) \\
\hline
$K_{\rm p}$ (\kms) &                        & $0.30^{+0.11}_{-0.08}$ &                        & $0.25^{+0.08}_{-0.06}$ & mod Jeffreys ($\sigma_{\rm RV}$) \\ 
$P_{\rm p}$ (d)    &                        & $36.16\pm0.11$         &                        & $28.19\pm0.10$         & Gaussian (36.2 or 28.2, 0.3) \\
T$_{\rm c}$ (2459000+) &                    & $731.9\pm1.8$          &                        & $734.7\pm1.7$          & Gaussian (732 or 735, 5) \\
$M_{\rm p} \sin i$ (\mjup) &                & $4.5^{+1.7}_{-1.2}$    &                        & $3.4^{+1.0}_{-0.8}$    & derived from $K_{\rm p}$, $P_{\rm p}$ and \mstar \\
\hline
\chisqr            & 2.08                   & 1.19                   & 0.89                   & 0.63                   & \\
rms (\kms)         & 0.24                   & 0.18                   & 0.21                   & 0.18                   & \\
$\log \mathcal{L}_M$& -14.3                 & -8.7                   & -7.6                   & 1.3                    & \\
$\log {\rm BF} = \Delta \log \mathcal{L}_M$ & 0.0     & 5.6          & 0.0                    & 8.9                    & \\
\hline
\end{tabular}
\label{tab:rv}
\end{table*}

\begin{figure*}
\centerline{\includegraphics[scale=0.6,angle=-90]{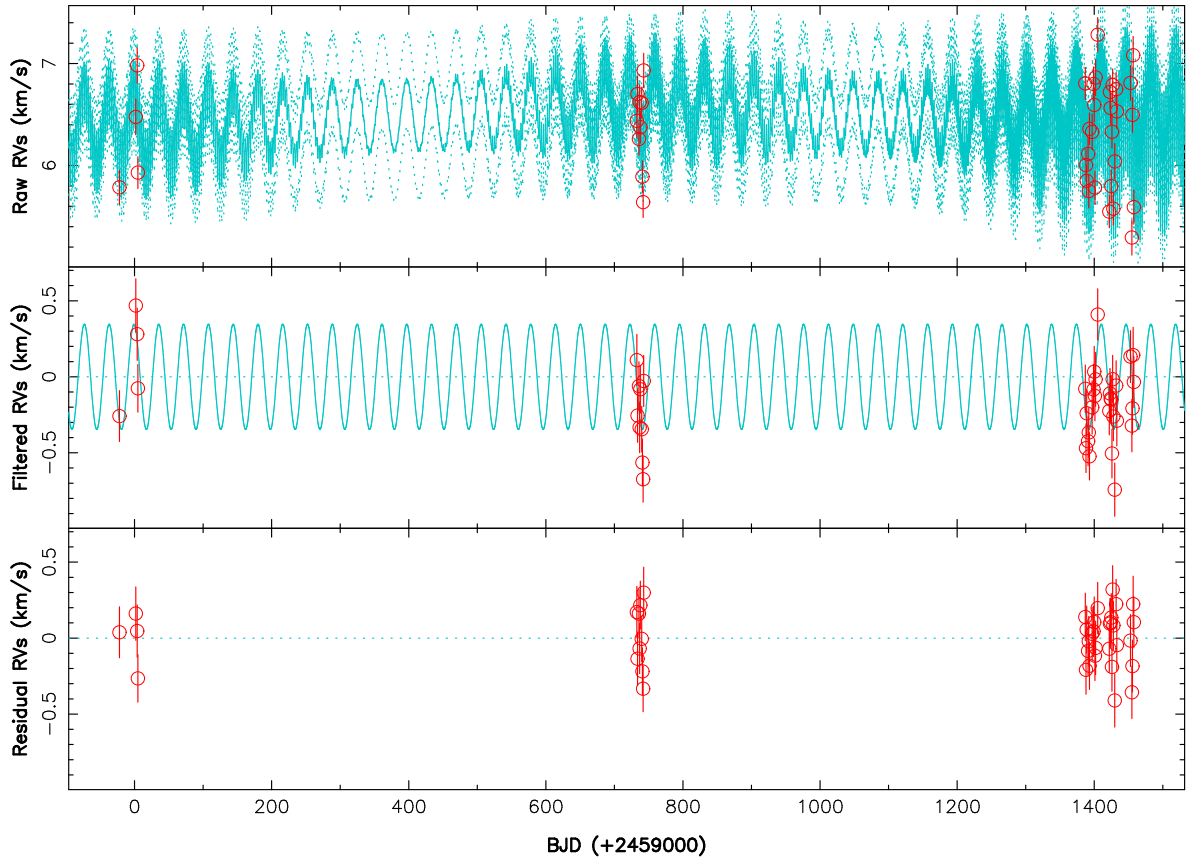}} 
\caption[]{Raw (top), activity filtered (middle) and residual (bottom) RVs derived from atomic lines of PDS~70 (red open circles).  The top plot shows the MCMC fit to the 
RV data, including a QP GPR modeling of the activity (cyan full line, with cyan dotted lines illustrating the 68~per cent confidence intervals), whereas the middle plot shows the 
RVs once activity is filtered out.  A zoom on our 2024 data is shown in Fig.~\ref{fig:rv24}.  Rotational modulation at \Prot\ dominates the dispersion of the raw RVs, 
especially in 2024 (see Fig.~\ref{fig:rv24}), whereas the additional medium-term fluctuation of the filtered RVs shown in the middle plot, also visible in the top plot, 
contributes to a smaller amount (see text).  The rms of the RV residuals is 0.18~\kms. } 
\label{fig:rv}
\end{figure*}

\begin{figure*}
\centerline{\includegraphics[scale=0.6,angle=-90]{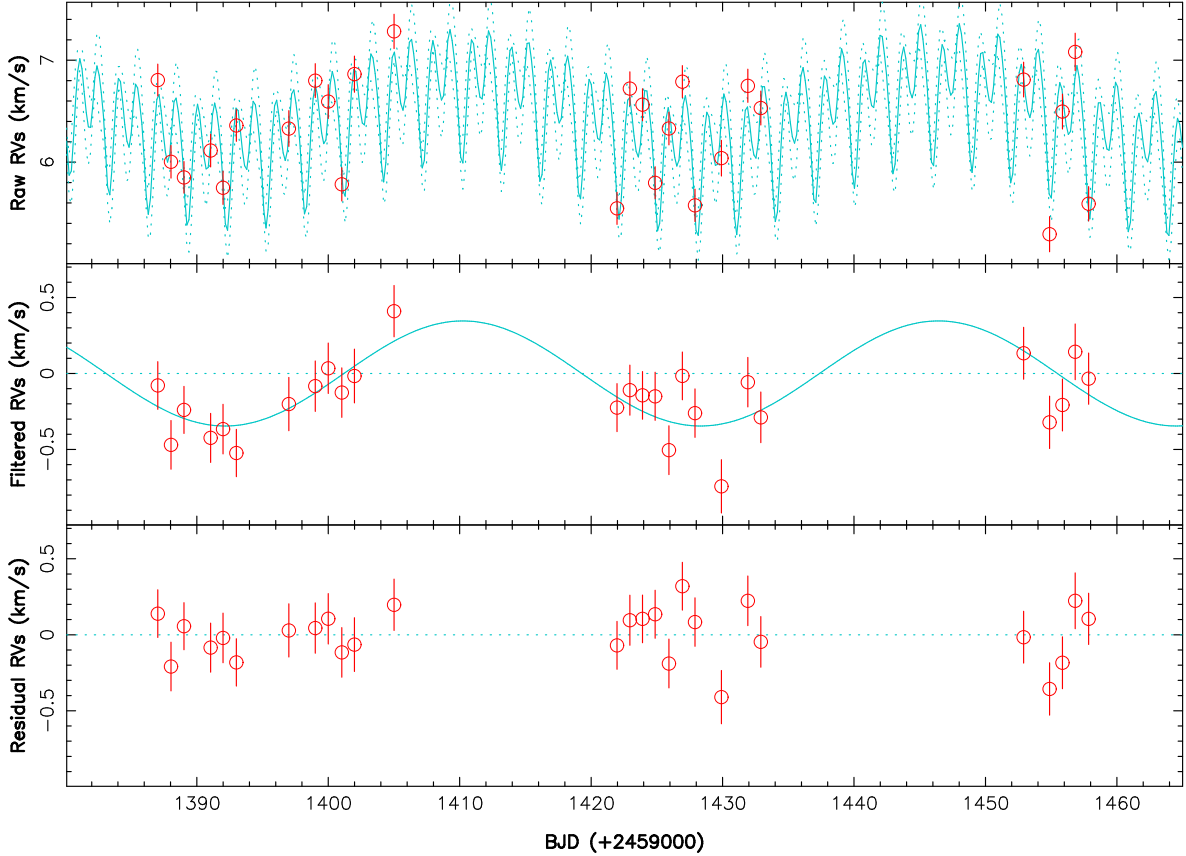}} 
\caption[]{Same as Fig.~\ref{fig:rv}, zooming on our 2024 data.}  
\label{fig:rv24}
\end{figure*}

\begin{figure}
\includegraphics[scale=0.48,angle=-90]{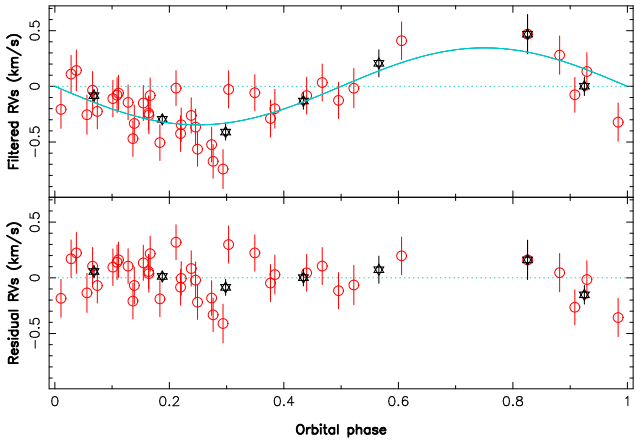}
\caption[]{Filtered (top plot) and residual (bottom plot) RVs of atomic lines of PDS~70 phase-folded on $P_{\rm p}$.  The red open circles are the individual RV points with 
the respective error bars, whereas the black stars are average RVs over 0.125 phase bins.  As in Fig.~\ref{fig:rv}, the dispersion of RV residuals is 0.18~\kms. 
Note the relative sparseness of points in the second half of the cycle, reflecting that $P_{\rm p}$ is close to one of the aliases of the synodic period 
of the Moon (SPIRou runs occurring in bright time). }
\label{fig:rvf}
\end{figure}

A simple periodogram of our RV data {\emr \citep[computed as in][]{Press92}} confirms that rotational modulation dominates the RV fluctuations, with the strongest peak located close to 
\Prot\ and associated with a false alarm probability (FAP) of $\simeq$0.01~per cent (see top plot of Fig.~\ref{fig:per} in Appendix~\ref{sec:appC}).  Fitting now our RVs with QP GPR 
(as previously achieved for our \Bl\ data, see Sec.~\ref{sec:bl}) confirms this result, yielding a period consistent with \Prot\ ($3.00\pm0.01$~d) and a GP amplitude 
$\theta_1=0.41^{+0.12}_{-0.09}$~\kms\ for the rotational modulation, as well as {\emr filtered RVs (i.e., the difference between the raw RVs and the derived GP)} whose rms of 0.24~\kms\ 
accounts for most of the excess RV dispersion (with respect to the median RV error bar of 0.16~\kms) but not for all ($\chisqr=2.08$, see Table~\ref{tab:rv}).  
As the periodogram of the filtered RVs suggests that power remains at periods in the range 20--40~d with a FAP level of $\simeq$2~per cent (see middle plot of Fig.~\ref{fig:per}),  
we repeated our QP GPR fitting, including this time an additional sine wave describing, e.g., the effect of a potential close-in planet in circular orbit around PDS~70, or that of 
another activity term fluctuating on a longer timescale than \Prot.  We detect an RV signal on top of rotational modulation, at a period of $P_{\rm p}=36.16\pm0.11$~d, with a semi-amplitude 
of $K_{\rm p}=0.30^{+0.11}_{-0.08}$~\kms\ (i.e. at a level slightly below 4~$\sigma$), which yields a significantly better fit to our RVs ($\chisqr=1.19$ and {\emr RV residuals, once both 
the GP and sinusoidal terms are subtracted,} of 0.18~\kms\ rms, see Table~\ref{tab:rv}).  In both cases, two GPR hyper-parameters, {\emr again weakly constrained by our sparse data (for 
the same reasons as in Sec.~\ref{sec:bl}), were fixed to a given value yielding maximum $\log \mathcal{L}_M$} in a preliminary run with all GPR variables free to vary, with virtually no 
difference on the outcome.  {\emr The selected value for $\theta_4$ is the same as that chosen for the modeling of \Bl\ though this parameter is usually smaller when adjusting RVs 
\citep[e.g.,][]{Donati23}, reflecting the limited precision (no better than 0.2) with which $\theta_4$ can be estimated in PDS~70.}  The corresponding increase in marginal log likelihood 
$\Delta \log \mathcal{L}_M = \log {\rm BF}$ reaches 5.6, suggesting that the detected medium-term RV signal is real \citep[following][]{Jeffreys61}.  The fit to the observed RVs as a 
function of time is depicted in Fig.~\ref{fig:rv} (with a zoom on the 2024 data in Fig.~\ref{fig:rv24}), whereas Fig.~\ref{fig:rvf} shows the filtered RVs phase-folded on $P_{\rm p}$.  
We note that the retrieved modulation period of the activity jitter is now slightly shorter than that reconstructed from \Bl\ data, again suggesting the presence of differential rotation 
at the surface of PDS~70.  This agrees with our ZDI results of Sec.~\ref{sec:zdi}, with RVs being more sensitive to equatorial features whereas \Bl\ apparently probes a larger latitude 
range and thereby reflects an average rotation rate at a higher latitude \citep[as in other TTSs, e.g.,][]{Donati24b}.  

Given our sparse data, the detected medium-term RV signal shows multiple aliases (e.g., at periods of 38.2, 34.3, and 40.3~d), but with logarithmic Bayes factors 
$\log {\rm BF}$ smaller by 0.5--1.0 than that of the peak signal at $P_{\rm p}$ (see Table~\ref{tab:rv}).  {\emr This signal is also present when considering the 2024 RV 
data alone, albeit with a lower significance ($\log {\rm BF}\simeq4$) and a larger error on $P_{\rm p}$ ($\simeq$5~d), as expected from the lower number of points and the shorter time span of our observations. }
If this RV signal is of planetary origin, the candidate planet 
would have a minimum mass of $M_{\rm p} \sin i=4.5^{+1.7}_{-1.2}$~\mjup, and hence a mass of $M_{\rm p} = 5.2^{+2.0}_{-1.4}$~\mjup\ if orbiting in the equatorial plane 
of the star.  Located at a distance of 0.2~au, it would orbit within, and close to the inner edge of, the inner disc of PDS~70.  
Interpreting this RV signal in terms of a planet is however premature, firstly because our data are too sparse 
(with only 40 points altogether collected over a period of 1480~d) for such an analysis to be conclusive, secondly because $P_{\rm p}$ is close to one of the distant 
aliases of the synodic period of the Moon in the window function (SPIRou runs occurring in bright time).  

{\emr To investigate this point in more depth, we looked at the bisector span (BIS) and the full width at half maximum (FWHM) of our Stokes $I$ LSD profiles of atomic lines (both listed in 
Table~\ref{tab:log}).  We find that BISs are clearly anti-correlated with RVs (Pearson's coefficient $R=-0.8$), confirming that RVs are primarily affected by rotational 
modulation.  A GPR fit to the BIS values (not shown) yields a dominant rotational modulation (with $\theta_1=0.8\pm0.2$~\kms\ and $\theta_2=3.000\pm0.008$~d), along with a 
lower amplitude signal ($K=0.3\pm0.1$~\kms) at one of the previously mentioned aliases ($40.3\pm0.1$~d) and a confidence level of $\log {\rm BF}=5.6$.  Performing again this 
analysis on FWHMs yields similar (though less precise) results, with a dominant rotational modulation ($\theta_1=0.6\pm0.2$~\kms\ and 
$\theta_2=3.02\pm0.04$~d), complemented by a weaker signal ($K=0.7\pm0.3$~\kms) of slightly lower significance ($\log {\rm BF}=4.7$) at 
yet another alias of the longer period ($32.3\pm0.1$~d).  Both findings cast further doubts on the potential planetary origin of the detected RV signal at 36~d, and rather 
suggest that it may be attributable to activity fluctuations on a timescale longer than \Prot\ (as those of \Bl\ and the parent large-scale field, see Secs.~\ref{sec:bl} 
and \ref{sec:zdi}).  Besides, we note that FWHMs are in average smaller in 2022 (by about 1.2~\kms) than in 2020 and 2024, indicating that the small-scale field (like the 
large-scale one, see Sec.~\ref{sec:zdi}) was weaker at this intermediate epoch of our observations.  We also find that RVs are not correlated with FWHMs, but rather to 
their first time derivative \citep[$R\simeq0.6$, as in, e.g.,][]{Donati23}. } 

{\emr To gather additional clues on the origin of this enigmatic RV signal,} we finally looked at the RVs derived from the Stokes $I$ LSD profiles of CO lines 
(see Sec.~\ref{sec:obs} and Table~\ref{tab:log}).  We find that these RVs are in average 0.45~\kms\ larger than those of atomic lines (convective downflows being relatively brighter 
for molecular than for atomic lines), and that they are less dispersed (with an rms of 0.34~\kms) as a result of CO lines being only sensitive to brightness features and not to magnetic 
fields.  We also infer that these RVs are 35~per cent less precise than those of atomic lines (with a median error bar of 0.22~\kms, see Table~\ref{tab:log}), due to the smaller number 
of lines used for LSD (only partly compensated by the CO lines being narrower than atomic lines).  We obtain that the activity jitter associated with rotational modulation is 
smaller for CO lines (with $\theta_1=0.26^{+0.10}_{-0.07}$~\kms) when fitting rotational modulation alone, yielding a residual rms of 0.21~\kms, and that an additional 
medium-term sinusoidal RV term is again needed to further improve $\log \mathcal{L}_M$ (by 8.9, see Table~\ref{tab:rv}).  However, the derived optimal period, now equal 
to $P_{\rm p}=28.19\pm0.11$~d, is hardly consistent with that obtained from the RVs, {\emr BISs and FWHMs} of atomic lines, and is even more 
suspicious, being quite close to one of the main 2-yr aliases of the synodic period of the Moon. 

Since the medium-term RV fluctuation detected in addition to rotational modulation in the LSD profiles of both atomic and CO bandhead lines is apparently 
chromatic, {\emr and is also present in the BISs and FWHMs of atomic lines}, 
we conclude that it is unlikely to be attributable to a candidate planet, but rather to activity at the surface of the star, e.g., caused by a change in the 
convective pattern resulting from the evolution of the large-scale magnetic field diagnosed in Secs.~\ref{sec:bl} and \ref{sec:zdi}, or within the inner disc regions.  
A similar dual-period magnetic modulation was recently detected in the \Bl\ curve of the young protostar V347~Aur \citep{Donati24c}.  Our study at least provides 
a conservative 3-$\sigma$ upper limit (of $\simeq$4~\mjup) on the minimum mass of a putative close-in giant planet in the inner disc of PDS~70, at a distance of about 0.2~au 
from the star.

\section{Accretion and wind}
\label{sec:eml}

To investigate accretion at the surface of PDS~70, we first looked at the spectral veiling, reducing the depth of spectral lines as a result of an added continuum 
(induced by accretion shocks at the surface of the star and by warm dust in the inner accretion disc, depending on wavelength).  Veiling is measured from the EWs of 
the Stokes $I$ LSD profiles of atomic lines ($r_{JH}$, in the $J$ and $H$ bands mostly) and of CO bandhead lines ($r_K$, in the $K$ band), comparing them with those obtained 
for a couple of non-accreting T~Tauri stars of similar spectral type (V819~Tau and TWA~9A) also observed with SPIRou.  We find that $r_{JH}$ is always smaller than 0.2, equal 
to $0.06\pm0.06$ in average, whereas $r_K$ is significantly larger, equal to $0.5\pm0.2$ in average and reaching up to 1.2 (see Table.~\ref{tab:log}).  Both are correlated 
($R\simeq0.6$) and neither of them exhibits rotational modulation, rather showing sporadic peaks at apparently random epochs.  Our estimates are consistent with those of 
\citet{Sousa23}, derived from a subset of the same SPIRou spectra (with a different method).

\begin{figure*}
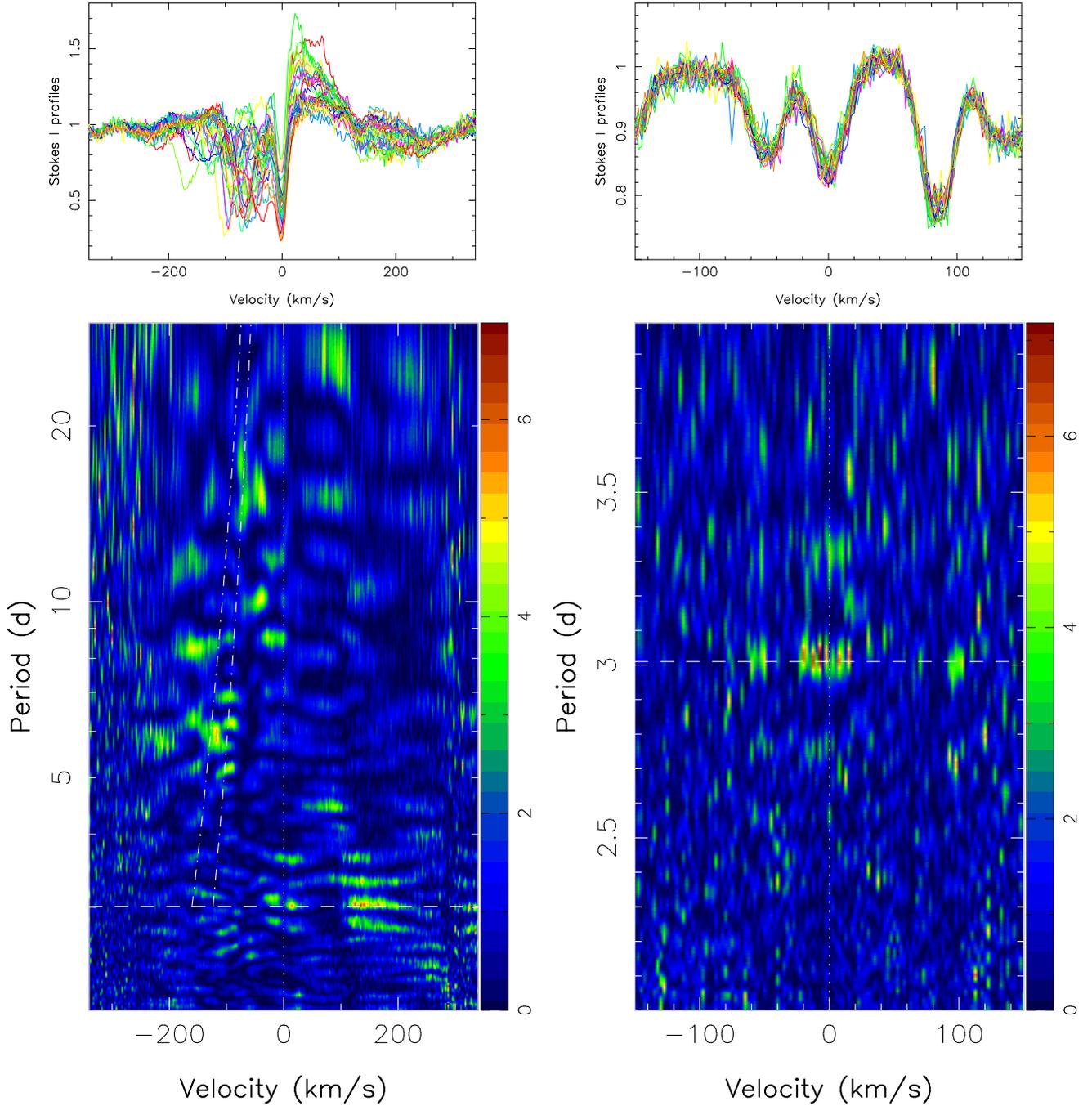

\centerline{\hspace{-2mm}\includegraphics[scale=0.3,angle=-90]{fig/pds70-hei24.ps}\hspace{19mm}\includegraphics[scale=0.3,angle=-90]{fig/pds70-pab24.ps}\vspace{2mm}}
\centerline{\includegraphics[scale=0.55,angle=-90]{fig/pds70-heiper24.ps}\hspace{3mm}\includegraphics[scale=0.55,angle=-90]{fig/pds70-pabper24.ps}}
\caption[]{Stacked Stokes $I$ profiles (top) and 2D periodograms (bottom) of the 1083.3-nm \hei\ triplet (left) and the 1282.16-nm \pab\ line (right), in the stellar 
rest frame, for our 2024 spectra of PDS~70.  In both periodograms, the dashed horizontal line traces \Prot.  The slanted dot-dash lines in the left panel illustrate how 
the Keplerian velocity and its line-of-sight projection vary as a function of the Keplerian period beyond \rcor\ in the inner disc.  The color scale depicts the logarithmic 
power. {\emr We caution that only the main peaks (colored yellow to red and extending over at least several velocity bins) are likely to be real in these plots.}   } 
\label{fig:eml1}
\end{figure*}

The 1083-nm \hei\ triplet is known to be an ideal proxy for investigating accretion and ejection flows of PMS stars and their surrounding discs, even those featuring 
low accretion rates like PDS~70 \citep{Thanathibodee20,Thanathibodee22,Campbell23}.  In PDS~70, both the blue-shifted absorption and the red-shifted emission components 
of the P-Cygni profile of the \hei\ triplet exhibit drastic variability in time and shape over our observation campaign.  In Fig.~\ref{fig:eml1} (top left panel), 
we show the stacked \hei\ profiles of our 2024 observations, featuring a highly variable blue-shifted absorption component extending at times down to $-200$~\kms, and 
a red-shifted component in emission up to about 125~\kms\ (in the stellar rest frame) and in absorption beyond this velocity.  We also detected a central narrow 
absorption component, putatively associated with chromospheric activity \citep{Thanathibodee22}, and previously observed in \citet{Campbell23} and \citet{Thanathibodee22} 
but not in \citet{Thanathibodee20}.  

The corresponding 2D periodogram (Fig.~\ref{fig:eml1}, bottom left panel) shows significant power at 3.03~d (and its 29.5-d aliases), i.e., a period slightly larger 
than that derived from \Bl\ and RVs (see Secs.~\ref{sec:bl} and \ref{sec:rvs}) and in the red-shifted absorption component only.  This red-shifted absorption presumably probes 
magnetospheric funnels regularly crossing the line of sight as the rotating star accretes material from the inner disc.  We note that the periodicity detected in this absorption 
component, again slightly larger than \Prot, is identical to that reported from the first TESS light curve \citep{Thanathibodee20, Gaidos24}, and likely relates to the rotation 
period of the high stellar latitudes at which accretion funnels are anchored, thus providing further evidence for differential rotation at the surface of PDS~70 (see 
Secs.~\ref{sec:zdi} and \ref{sec:rvs}). Besides, both the red and blue wings of the central chromospheric absorption are also modulated with \Prot, suggesting an RV 
variation of this contribution potentially tracing a low-latitude chromospheric region at the surface of the star.  No Zeeman signature is detected in conjunction with this 
central component, confirming that it is unlikely related to the high-latitude footpoints of funnel flows linking the star to the inner edge of the disc \citep[conversely 
to what was found on, e.g., TW~Hya,][]{Donati24b}.  

The blue-shifted absorption, likely probing a wind from the inner disc \citep{Thanathibodee22,Campbell23} given its shape and temporal variability \citep{Edwards03,Kwan07}, exhibits 
distributed power at periods longer than \Prot, ranging from from 5 to 20~d (see Fig.~\ref{fig:eml1}) and corresponding to Keplerian periods at distances of 0.055 to 0.14~au.  
We speculate that these features reflect azimuthal structures in the inner disc wind of PDS~70, possibly probing magnetically active regions where the wind is stronger and 
potentially linked to the magnetically driven wind traced with the 630-nm \foxi\ \citep[originating from the same region of the inner disc,][]{Campbell23}.  
We note a potential trend for longer periods to show up at smaller blue shifts, suggesting disc wind outward velocities scaling with Keplerian velocities in the disc 
\citep{Blandford82, Thanathibodee22}, decreasing from 135 to 85~\kms\ for periods in the range 5 to 20~d in PDS~70.  Finally, we see no clear periodic modulation in the red-shifted 
emission component between 0 and 120~\kms, presumably probing the top part of accretion funnels close to \rcor.  This suggests that accretion flows from the inner disc to 
the star are likely to be unsteady, i.e., irregular from one rotation cycle to the next, as red-shifted emission would otherwise be expected to exhibit some level of rotational 
modulation as accretion columns rotate around the star.  

\begin{figure}
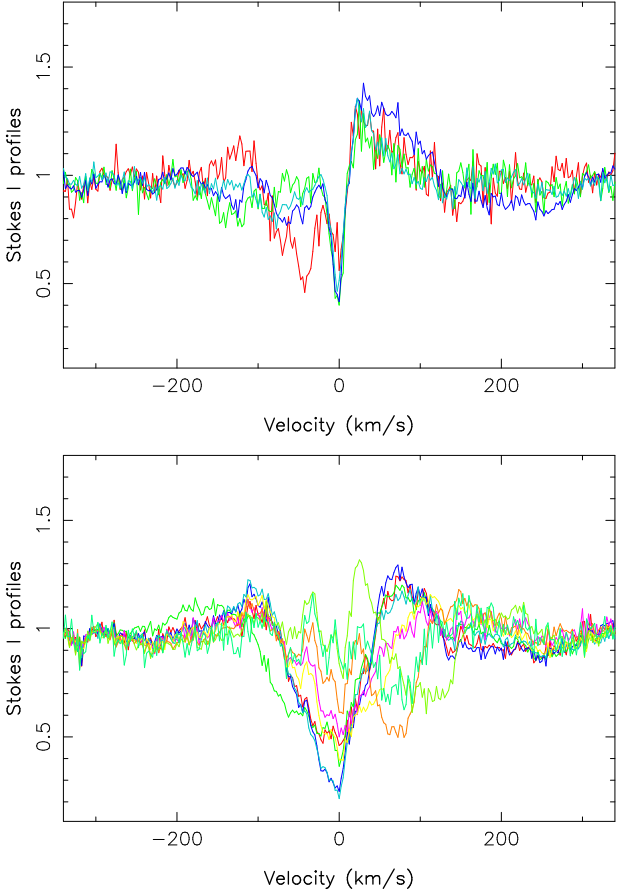

\includegraphics[scale=0.35,angle=-90]{fig/pds70-hei20.ps}\vspace{2mm}
\includegraphics[scale=0.35,angle=-90]{fig/pds70-hei22.ps}
\caption[]{Same as top left panel \ref{fig:eml1}, for our 2020 (top) and 2022 (bottom) \hei\ spectra of PDS~70.  The 3 profiles in the bottom panel showing strong red-shifted 
absorption at velocities 50$-$150~\kms\ correspond to the last 3 spectra collected in June 2022.  } 
\label{fig:eml2}
\end{figure}

We also present in Fig.~\ref{fig:eml2} the \hei\ triplet profiles of PDS~70 during our 2020 and 2022 observations, showing a behaviour similar to that of our 2024 
observations with the exception of the red-shifted emission component up to 125~\kms\ that switched to absorption in the last 3 spectra collected in June 2022, as 
described in previous studies \citet{Thanathibodee20,Thanathibodee22}.  This supports our previous conclusion that accretion flows from the inner disc are unsteady in 
PDS~70, featuring sporadic episodes of stronger absorption from the inner disc like the one which occurred during our last 3 observations of June 2022.  We stress that this 
episode of enhanced accretion occurred at an epoch where the large-scale field of PDS~70 is significantly smaller than average (see Sec.~\ref{sec:zdi}).  

We find that \pab\ in PDS~70 shows very little temporal variability (see top right panel of Fig.~\ref{fig:eml1} for our 2024 observations).  The corresponding 2D 
periodogram (bottom right panel) shows that this variability is modulated with \Prot\ and confined to a velocity range of $\pm\vsini$ (as for the neighbouring 
\cai\ and \tii\ lines on the left and the right of \pab\ respectively, but with significantly more power).  This variability is most likely of chromospheric origin, 
as for the central absorption component of \hei.  The modulated component of \pab\ amounts to a peak-to-peak EW variation of about 0.4~\kms\ (1.7~pm).  The \pab\ 
profiles of PDS~70 in our 2020 and 2022 observations were very similar to those in 2024, with variations from epoch to epoch comparable to the modulation detected in 
2024.  Besides, we find that the EW of \pab\ in PDS~70 is equal to $5.6\pm0.4$~\kms\ ($24\pm2$~pm) at all epochs (even in our last 3 spectra of June 2022), i.e., 
identical within error bars to that in our comparison weak-line TTSs V819~Tau and TWA~9A.  Using the scaling laws reported between emission line fluxes and accretion 
luminosities of accreting TTSs \citep{Alcala17}, this would imply a 3-$\sigma$ upper limit on $\log\Mdot$ of $-11$ (in \mspy) at the surface of PDS~70, much lower than the 
reported values \citep[of order $-10$,][]{Thanathibodee20,Campbell23,Gaidos24} that should imply \pab\ EW differences (with respect to non-accreting TTSs) and / or temporal 
variations of order 4~\kms\ (20~pm), i.e., 10$\times$ larger than those we observed.  That we do not see excess \pab\ flux beyond velocities of $\pm$\vsini\ is further 
evidence that whichever weak mass accretion rate is occurring on PDS~70 is undetected in the \pab\ profiles of our SPIRou spectra despite the relatively high SNR.  
The same conclusion holds for \brg, for which no variability (from either chromospheric activity or accretion) is observed within the line profile nor beyond 
velocities of $\pm$\vsini, and with an average (veiling corrected) EW of $1.8\pm0.5$~\kms\ ($13\pm4$~pm) in PDS~70 that is only marginally different (by 
$0.8\pm0.5$~\kms\ or $6\pm4$~pm) from that in our comparison non-accreting TTSs.  It yields again a 3-$\sigma$ upper limit on $\log\Mdot$ of $-11$ (in \mspy) at the 
surface of PDS~70 using the scaling laws of \citet{Alcala17}.  
A potential explanation for this discrepancy is that PDS~70 was in a state of particularly low accretion rate at the various epochs of our observations.  More likely, 
we suspect that the scaling laws of \citet{Alcala17}, whose average precision is about a factor of 3 on accretion luminosities derived from \pab\ and \brg\ fluxes, may 
be less accurate or more dispersed for logarithmic accretion luminosities (relative to \lsun) in the range $-3$ to $-4$ of relevance for PDS~70, being based on a 
comparatively larger number of upper limits than for larger accretion luminosities.   

Using the previously published logarithmic mass accretion rate along with our estimates of the dipole component of the large-scale field of PDS~70 (see 
Sec.~\ref{sec:zdi}), we computed the distance \rmag\ at which the field is able to disrupt the inner disc, using simulation results of magnetospheric accretion 
\citep{Zanni13,Blinova16,Blinova19,Pantolmos20} and in particular the analytical expression of \citet{Bessolaz08} based on such simulations and confirmed in newer 
ones.  For $\log\Mdot$ between $-9.7$ and $-10.3$ (in units of \mspy, see Table~\ref{tab:par}), we find that \rmag\ ranges from 5.6 to 8.3~\rstar\ for the strongest 
dipole field we measured on PDS~70 (420~G, see Table~\ref{tab:mag}), and from 3.7 to 5.4~\rstar\ for the weakest dipole field (200~G).  This implies 
$\rmag/\rcor=1.08\pm0.21$ in the strong field case and $\rmag/\rcor=0.70\pm0.14$ in the weak field one, or $\rmag/\rcor=0.92\pm0.18$ for an intermediate case 
(corresponding to 320~G, i.e., the average dipole field of PDS~70).  
It confirms in particular that the inner edge of the inner disc of PDS~70 is in average located close to \rcor, moving towards the star at epochs of weaker 
dipole field (as in 2022 June) or extending beyond \rcor\ at epochs of stronger dipole field (as in 2024 April or 2020 May-June).  Whereas most disc material is expected to 
be accreted onto the star in the weaker field case, a magnetic propeller mechanism \citep{Romanova04,Ustyugova06} can be triggered in the strong field case  
where a smaller amount of disc material is accreted onto the star, leading to variations of $\log\Mdot$ and sporadic eruptive episodes \citep{DAngelo12,Zanni13}.

\section{Summary, discussion and conclusion}
\label{sec:dis}

In this paper we report spectropolarimetric and velocimetric observations of the weakly accreting, planet hosting T~Tauri star PDS~70, collected with SPIRou at CFHT 
in 2020 May-June, 2022 June and 2024 March-May.  We detect clear Zeeman signatures in the LSD Stokes $V$ profiles of PDS~70, implying longitudinal magnetic fields \Bl\ 
ranging from $-116$ to 176~G and modulated with the average stellar rotation period of $\Prot=3.008\pm0.006$~d, consistent with the recurrence period derived from the 
archival ASAS-SN $V$-band and $g$-band light curves of PDS~70.  Splitting our LSD Stokes $I$ and $V$ profiles into 5 subsets corresponding to different epochs (3 of which 
in 2024), we reconstructed the brightness and magnetic maps of PDS~70 and their changes with time.  We find that PDS~70 hosts a large-scale magnetic field that evolves with 
time on a timescale of a few months, with a dominant dipole component varying from 200 (in 2022) to about 400~G (in 2020 and 2024), with an average of 320~G.  
Our 2024 observations, spread over 3 months, show clear 
month-to-month changes in the large-scale field topology, both in the orientation and strength of the dipole field.  The magnetic topology of PDS~70 is consistent 
with those of partly-convective TTSs, often showing weaker and more rapidly varying fields than fully-convective ones \citep{Donati09}.  From the broadening of spectral 
lines in our median spectrum, we find that PDS~70 hosts a small-scale field of $2.51\pm0.12$~kG, typical to that of TTSs \citep[e.g.,][]{Johns07,Lopez-Valdivia21}.  
{\emr We also infer that FWHMs of atomic lines are smallest in 2022, indicating that the small-scale field of PDS~70 was weaker at this intermediate epoch than in 2020 and 2024}.  

Our observations also suggest that the surface of PDS~70 is likely sheared by latitudinal differential rotation, with the pole rotating more slowly than the equator. 
RVs, more sensitive to low-latitude surface inhomogeneities, indicate recurrence periods down to 2.99~d  whereas the red-shifted absorption component of the \hei\ line, 
probing magnetospheric funnels anchored at high stellar latitudes, points to a longer rotation period of 3.03~d.  It implies a level of differential rotation between the 
low and high latitudes of $\simeq$28~\mrpd, about half that at the surface of the Sun.  This is again typical to non-fully convective stars, often exhibiting differential 
rotation rates larger than fully convective ones, even at rotation rates much larger than that of the Sun \citep[e.g.,][]{Donati09}.  

We also investigated whether PDS~70 hosts a close-in giant planet in addition to the two distant ones already revealed by direct imaging \citep{Haffert19}.  We find that 
RVs derived from atomic lines of PDS~70 exhibit variability at an rms level of 0.49~\kms, featuring time-dependant rotational modulation with an average semi-amplitude of 
0.44~\kms\ as a result of brightness and magnetic regions at the surface of the star.  Besides, we show that RVs from the CO bandhead lines in the $K$ band, insensitive to 
magnetic fields, are also rotationally modulated with a twice weaker amplitude, reflecting the impact of surface brightness inhomogeneities 
on the RV curve of PDS~70.  RVs from both atomic and CO lines are apparently also fluctuating on a longer timescale of about 28$-$36~d (with semi-amplitudes of 0.30 and 
0.25~\kms\ respectively), but most likely as a result of activity rather than of a putative close-in giant planet around PDS~70 {\emr given that BISs and FWHMs of LSD 
Stokes $I$ profiles of atomic lines are also showing a similar variation pattern}.  We derive a 3-$\sigma$ upper limit of 
$\simeq$4~\mjup\ for the minimum mass of a close-in planet at a distance of about 0.2~au.  The physical nature of this activity is currently unclear.  Whereas the period 
on which it fluctuates may suggest it originates from the inner disc at a distance of $\simeq$0.2~au, i.e., in a region from which a magnetically driven wind is launched 
\citep{Campbell23}, the RV fluctuation it generates (of semi-amplitude 0.25$-$0.30~\kms) rather argues in favour of a stellar phenomenon (e.g., granulation) given both the 
low density of the inner disc and the magnitude of the spectral contribution required to induce such a perturbation.  

We examined the usual accretion proxies and their temporal variations in the spectrum of PDS~70, and in particular the 1083.3-nm \hei\ triplet of PDS~70 known to be a very 
sensitive probe of magnetospheric accretion towards the host star and of magnetically driven winds from the inner accretion disc.  We find that the \hei\ profiles of PDS~70 
show both blue-shifted and red-shifted extended absorption, bracketing a red-shifted emission component and a narrow central (likely chromospheric) absorption feature.  
Only the extended red-shifted and the central chromospheric components exhibit rotational modulation, the former likely reflecting accretion funnels periodically crossing the 
line of sight as PDS~70 rotates.  This modulation of the red-shifted \hei\ absorption component confirms that accretion still takes place at the surface of PDS~70.  
Power at periods ranging from 5 to 20~d is detected in the blue-shifted absorption component at velocities ranging down to $-150$~\kms, potentially tracing azimuthal 
structures in the inner disc wind of PDS~70, e.g., from magnetic regions of the disc where the wind is stronger.  These structures may be the anchor points of the magnetically 
driven wind traced with the 630-nm \foxi\ \citep[originating from the same region of the inner disc,][]{Campbell23}.

Only chromospheric activity (inducing rotational modulation of the line EW at a level of a few pm within velocites of $\pm$\vsini) but no sign of accretion (within 
and beyond $\pm$\vsini) is detected in \pab\ down to EWs of $\pm$0.4~\kms, whereas \brg\ shows no clear sign of either.  Using the scaling relations linking \pab\ 
and \brg\ line fluxes to accretion luminosities \citep[e.g.,][]{Alcala17}, we infer a 3-$\sigma$ upper limit on $\log\Mdot$ of $-11$ (in \mspy) at the surface of PDS~70.  
It either suggests that PDS~70 was in a state of unusually low accretion rate during all of our observations, or more realistically, that the scaling relations of 
\citet[][]{Alcala17}, yielding accretion luminosities from \pab\ and \brg\ line fluxes, may be less accurate and more dispersed at the low accretion rates reported for PDS~70.  

Given the published accretion rate and the large-scale magnetic topology derived from our study, we can conclude that PDS~70 is apparently able to disrupt the inner 
accretion disc up to an average distance (relative to \rcor) of $\rmag/\rcor=0.92\pm0.18$ (the error bar reflecting the reported range of $\log\Mdot$), with extreme values 
spanning from $0.70\pm0.14$ to $1.08\pm0.21$ at epochs of weakest and strongest dipole field components respectively.  This variation of the large-scale field can in 
particular qualitatively explain fluctuations in the accretion rate at the surface of PDS~70, with most of the disc material being accreted onto the star through conventional 
magnetospheric accretion patterns when the dipole field is on the weak side (as in 2022 June), or only a small fraction of it reaching the stellar surface when the 
dipole field is strongest and the accretion mode switches to a magnetic propeller regime \citep{Romanova04,Ustyugova06,DAngelo12,Zanni13}.  

Comparing with the ASAS-SN $g$-band light curve (see Fig.~\ref{fig:lc2}), we find that PDS~70 is in average fainter when the small-scale field and the dipole component of the 
large-scale magnetic field are weaker, i.e., in 2022 June, also coinciding with PDS~70 being brighter at 3.4 and 4.6~\mic\ \citep{Gaidos24}.  These photometric changes apparently 
result from occultation by sub-\mic\ dust from the inner disc dimming the star at optical wavelengths and generating additional 3$-$5~\mic\ emission \citep{Gaidos24}.  This leads 
us to speculate that dust is likely able to pile up at \rcor\ when the dipole component of the large-scale field is strong enough to ensure that \rmag\ extends up \rcor\ 
\citep[as proposed by][]{Sanderson23}, like in our 2020 and 2024 SPIRou observations.  In such cases, PDS~70 suffers less occultation and less reddening, with dust clouds 
located at \rcor\ in the disc plane only marginally intersecting the line of sight given the viewing angle, and the optical light curves tends to be periodic and reflect mostly 
stellar activity.  As dust clouds 
at \rcor\ also rotate with \Prot\ and are stable against gravity, the light curve can at times feature a ``scallop-shell'' morphology like that collected by TESS in sector 11 
when dust clouds happen to occult a larger fraction of the stellar disc.  
On the opposite, when the dipole component of PDS~70 gets weaker and \rmag\ is significantly smaller than \rcor, dust is able to penetrate the magnetosphere and to occult 
the stellar disc more substantially, generating lower average optical fluxes and larger-amplitude stochastic light curves like those collected by TESS in sectors 38 and 65 
as dust within the magnetosphere rapidly drifts towards the star being no longer supported against gravity.  Our scenario is at least able to qualitatively reconcile our 
observations with those of \citet{Gaidos24}, despite their conclusions differing from ours on how magnetic fields drive the process.  

Our study demonstrates the key value of repeated spectropolarimetric and velocimetric observations of the PMS planet-hosting star PDS~70 carried out over multiple 
seasons, like those we secured with SPIRou, to further characterize the central star and the inner accretion disc, the magnetospheric accretion processes linking 
both and the wind escaping from the disc, as well as any planet that may hide within 1~au of the host star.  
With its two outer giant planets, PDS~70 is an ideal target to investigate whether more such planets can also be present at smaller orbital distances, 
e.g., after undergoing inward migration over the last few Myr since their initial formation beyond the snow line.  Even though we were only able to obtain an upper 
limit on the mass of such a putative close-in giant planet at 0.2~au with the present study, regular observations carried out over a longer timespan of about a decade should 
be able to reach a more definite conclusion on this point up to distances of a few au's, and provide at the same time essential information on the temporal evolution of 
the large-scale magnetic field of the host 
star, on how this evolving field impacts the way material is accreted from the disc, and on the properties of the magnetically-driven wind escaping from the central 
disc regions.  We thus strongly advocate for renewed SPIRou observations of PDS~70 in forthcoming seasons, in particular at epochs where complementary data on a wider 
wavelength domain can be collected (e.g., with ASAS-SN, TESS, ESPaDOnS at CFHT, and HARPS and X-Shooter at ESO).

\section*{Acknowledgements}
We thank an anonymous referee for valuable comments on an earlier version of our manuscript.  
This project received funds from the European Research Council (ERC) under the H2020 and Horizon Europe research \& innovation programs (grant agreements \#740651 NewWorlds, 
\#742095 SPIDI, \#101053020  Dust2Planets, \#101039452 WANDA),  
from the Agence Nationale pour la Recherche (ANR, project ANR-18-CE31-0019 SPlaSH) and from the Investissements d'Avenir program (project ANR-15-IDEX-02), 
and was also supported by the NKFIH excellence grant TKP2021-NKTA-64.  SHPA acknowledges financial support from CNPq, CAPES and Fapemig.  
This work benefited from the SIMBAD CDS database at URL {\tt http://simbad.u-strasbg.fr/simbad}, the ADS system at URL {\tt https://ui.adsabs.harvard.edu} 
and the ASAS-SN archival photometry {\tt http://asas-sn.osu.edu}. 
Our study is based on data obtained at the CFHT, operated by the CNRC (Canada), INSU/CNRS (France) and the University of Hawaii.
The authors wish to recognise and acknowledge the very significant cultural role and reverence that the summit of Maunakea has always had
within the indigenous Hawaiian community.  We are most fortunate to have the opportunity to conduct observations from this mountain.

\section*{Data availability}  SLS data are publicly available from the Canadian Astronomy Data Center, whereas SPICE data will be available at the same place 
by mid 2025.

\bibliography{pds70} 

\begin{thebibliography}{}
\makeatletter
\relax
\def\mn@urlcharsother{\let\do\@makeother \do\$\do\&\do\#\do\^\do\_\do\%\do\~}
\def\mn@doi{\begingroup\mn@urlcharsother \@ifnextchar [ {\mn@doi@}
  {\mn@doi@[]}}
\def\mn@doi@[#1]#2{\def\@tempa{#1}\ifx\@tempa\@empty \href
  {http://dx.doi.org/#2} {doi:#2}\else \href {http://dx.doi.org/#2} {#1}\fi
  \endgroup}
\def\mn@eprint#1#2{\mn@eprint@#1:#2::\@nil}
\def\mn@eprint@arXiv#1{\href {http://arxiv.org/abs/#1} {{\tt arXiv:#1}}}
\def\mn@eprint@dblp#1{\href {http://dblp.uni-trier.de/rec/bibtex/#1.xml}
  {dblp:#1}}
\def\mn@eprint@#1:#2:#3:#4\@nil{\def\@tempa {#1}\def\@tempb {#2}\def\@tempc
  {#3}\ifx \@tempc \@empty \let \@tempc \@tempb \let \@tempb \@tempa \fi \ifx
  \@tempb \@empty \def\@tempb {arXiv}\fi \@ifundefined
  {mn@eprint@\@tempb}{\@tempb:\@tempc}{\expandafter \expandafter \csname
  mn@eprint@\@tempb\endcsname \expandafter{\@tempc}}}

\bibitem[\protect\citeauthoryear{{Alcal{\'a}} et~al.,}{{Alcal{\'a}}
  et~al.}{2017}]{Alcala17}
{Alcal{\'a}} J.~M.,  et~al., 2017, \mn@doi [\aap]
  {10.1051/0004-6361/201629929}, \href
  {https://ui.adsabs.harvard.edu/abs/2017A&A...600A..20A} {600, A20}

\bibitem[\protect\citeauthoryear{{Andr{\'e}}, {Di Francesco}, {Ward-Thompson},
  {Inutsuka}, {Pudritz}  \& {Pineda}}{{Andr{\'e}} et~al.}{2014}]{Andre14}
{Andr{\'e}} P.,  {Di Francesco} J.,  {Ward-Thompson} D.,  {Inutsuka} S.~I.,
  {Pudritz} R.~E.,   {Pineda} J.~E.,  2014, in {Beuther} H.,  {Klessen} R.~S.,
  {Dullemond} C.~P.,   {Henning} T.,  eds, Protostars and Planets VI. pp 27--51
  (\mn@eprint {arXiv} {1312.6232}),
  \mn@doi{10.2458/azu_uapress_9780816531240-ch002}

\bibitem[\protect\citeauthoryear{{Artigau} et~al.,}{{Artigau}
  et~al.}{2022}]{Artigau22}
{Artigau} {\'E}.,  et~al., 2022, \mn@doi [\aj] {10.3847/1538-3881/ac7ce6},
  \href {https://ui.adsabs.harvard.edu/abs/2022AJ....164...84A} {164, 84}

\bibitem[\protect\citeauthoryear{{Baraffe}, {Homeier}, {Allard}  \&
  {Chabrier}}{{Baraffe} et~al.}{2015}]{Baraffe15}
{Baraffe} I.,  {Homeier} D.,  {Allard} F.,   {Chabrier} G.,  2015, \mn@doi
  [\aap] {10.1051/0004-6361/201425481}, \href
  {http://adsabs.harvard.edu/abs/2015A%26A...577A..42B} {577, A42}

\bibitem[\protect\citeauthoryear{{Benisty} et~al.,}{{Benisty}
  et~al.}{2021}]{Benisty21}
{Benisty} M.,  et~al., 2021, \mn@doi [\apjl] {10.3847/2041-8213/ac0f83}, \href
  {https://ui.adsabs.harvard.edu/abs/2021ApJ...916L...2B} {916, L2}

\bibitem[\protect\citeauthoryear{{Bessolaz}, {Zanni}, {Ferreira}, {Keppens}  \&
  {Bouvier}}{{Bessolaz} et~al.}{2008}]{Bessolaz08}
{Bessolaz} N.,  {Zanni} C.,  {Ferreira} J.,  {Keppens} R.,   {Bouvier} J.,
  2008, \mn@doi [\aap] {10.1051/0004-6361:20078328}, \href
  {http://adsabs.harvard.edu/abs/2008A%26A...478..155B} {478, 155}

\bibitem[\protect\citeauthoryear{{Blakely} et~al.,}{{Blakely}
  et~al.}{2024}]{Blakely24}
{Blakely} D.,  et~al., 2024, \mn@doi [arXiv e-prints]
  {10.48550/arXiv.2404.13032}, \href
  {https://ui.adsabs.harvard.edu/abs/2024arXiv240413032B} {p. arXiv:2404.13032}

\bibitem[\protect\citeauthoryear{{Blandford} \& {Payne}}{{Blandford} \&
  {Payne}}{1982}]{Blandford82}
{Blandford} R.~D.,  {Payne} D.~G.,  1982, \mn@doi [\mnras]
  {10.1093/mnras/199.4.883}, \href
  {https://ui.adsabs.harvard.edu/abs/1982MNRAS.199..883B} {199, 883}

\bibitem[\protect\citeauthoryear{{Blinova}, {Romanova}  \&
  {Lovelace}}{{Blinova} et~al.}{2016}]{Blinova16}
{Blinova} A.~A.,  {Romanova} M.~M.,   {Lovelace} R.~V.~E.,  2016, \mn@doi
  [\mnras] {10.1093/mnras/stw786}, \href
  {http://adsabs.harvard.edu/abs/2016MNRAS.459.2354B} {459, 2354}

\bibitem[\protect\citeauthoryear{{Blinova}, {Romanova}, {Ustyugova}, {Koldoba}
  \& {Lovelace}}{{Blinova} et~al.}{2019}]{Blinova19}
{Blinova} A.~A.,  {Romanova} M.~M.,  {Ustyugova} G.~V.,  {Koldoba} A.~V.,
  {Lovelace} R.~V.~E.,  2019, \mn@doi [\mnras] {10.1093/mnras/stz1314}, \href
  {https://ui.adsabs.harvard.edu/abs/2019MNRAS.487.1754B} {487, 1754}

\bibitem[\protect\citeauthoryear{{Bouvier}}{{Bouvier}}{2022}]{Bouvier22}
{Bouvier} J.,  2022, in The 21st Cambridge Workshop on Cool Stars, Stellar
  Systems, and the Sun. Cambridge Workshop on Cool Stars, Stellar Systems, and
  the Sun.
p.~40, \mn@doi{10.5281/zenodo.7373044}

\bibitem[\protect\citeauthoryear{{Braun}, {Yen}, {Koch}, {Manara}, {Miotello}
  \& {Testi}}{{Braun} et~al.}{2021}]{Braun21}
{Braun} T. A.~M.,  {Yen} H.-W.,  {Koch} P.~M.,  {Manara} C.~F.,  {Miotello} A.,
    {Testi} L.,  2021, \mn@doi [\apj] {10.3847/1538-4357/abd24f}, \href
  {https://ui.adsabs.harvard.edu/abs/2021ApJ...908...46B} {908, 46}

\bibitem[\protect\citeauthoryear{{Brown}, {Donati}, {Rees}  \& {Semel}}{{Brown}
  et~al.}{1991}]{Brown91}
{Brown} S.~F.,  {Donati} J.-F.,  {Rees} D.~E.,   {Semel} M.,  1991, \aap, \href
  {http://adsabs.harvard.edu/abs/1991A%26A...250..463B} {250, 463}

\bibitem[\protect\citeauthoryear{{Cabrit}}{{Cabrit}}{2024}]{Cabrit24}
{Cabrit} S.,  2024, in EAS2024. p.~2674

\bibitem[\protect\citeauthoryear{{Campbell-White} et~al.,}{{Campbell-White}
  et~al.}{2023}]{Campbell23}
{Campbell-White} J.,  et~al., 2023, \mn@doi [\apj] {10.3847/1538-4357/acf0c0},
  \href {https://ui.adsabs.harvard.edu/abs/2023ApJ...956...25C} {956, 25}

\bibitem[\protect\citeauthoryear{{Chambers}}{{Chambers}}{2018}]{Chambers18}
{Chambers} J.,  2018, \mn@doi [\apj] {10.3847/1538-4357/aada09}, \href
  {https://ui.adsabs.harvard.edu/abs/2018ApJ...865...30C} {865, 30}

\bibitem[\protect\citeauthoryear{{Chib} \& {Jeliazkov}}{{Chib} \&
  {Jeliazkov}}{2001}]{Chib01}
{Chib} S.,  {Jeliazkov} I.,  2001, Journal of the American Statistical
  Association, \href {http://adsabs.harvard.edu/abs/2016arXiv160800962B} {96,
  270}

\bibitem[\protect\citeauthoryear{{Claret}, {Diaz-Cordoves}  \&
  {Gimenez}}{{Claret} et~al.}{1995}]{Claret95}
{Claret} A.,  {Diaz-Cordoves} J.,   {Gimenez} A.,  1995, \aaps, \href
  {https://ui.adsabs.harvard.edu/abs/1995A&AS..114..247C} {114, 247}

\bibitem[\protect\citeauthoryear{{Cristofari} et~al.,}{{Cristofari}
  et~al.}{2023a}]{Cristofari23}
{Cristofari} P.~I.,  et~al., 2023a, \mn@doi [\mnras] {10.1093/mnras/stad865},
  \href {https://ui.adsabs.harvard.edu/abs/2023MNRAS.522.1342C} {522, 1342}

\bibitem[\protect\citeauthoryear{{Cristofari} et~al.,}{{Cristofari}
  et~al.}{2023b}]{Cristofari23b}
{Cristofari} P.~I.,  et~al., 2023b, \mn@doi [\mnras] {10.1093/mnras/stad3144},
  \href {https://ui.adsabs.harvard.edu/abs/2023MNRAS.526.5648C} {526, 5648}

\bibitem[\protect\citeauthoryear{{D'Angelo} \& {Spruit}}{{D'Angelo} \&
  {Spruit}}{2012}]{DAngelo12}
{D'Angelo} C.~R.,  {Spruit} H.~C.,  2012, \mn@doi [\mnras]
  {10.1111/j.1365-2966.2011.20046.x}, \href
  {https://ui.adsabs.harvard.edu/abs/2012MNRAS.420..416D} {420, 416}

\bibitem[\protect\citeauthoryear{{David}, {Petigura}, {Luger},
  {Foreman-Mackey}, {Livingston}, {Mamajek}  \& {Hillenbrand}}{{David}
  et~al.}{2019}]{David19}
{David} T.~J.,  {Petigura} E.~A.,  {Luger} R.,  {Foreman-Mackey} D.,
  {Livingston} J.~H.,  {Mamajek} E.~E.,   {Hillenbrand} L.~A.,  2019, \mn@doi
  [\apjl] {10.3847/2041-8213/ab4c99}, \href
  {https://ui.adsabs.harvard.edu/abs/2019ApJ...885L..12D} {885, L12}

\bibitem[\protect\citeauthoryear{{Donati} \& {Brown}}{{Donati} \&
  {Brown}}{1997}]{Donati97c}
{Donati} J.-F.,  {Brown} S.~F.,  1997, \aap, \href
  {http://adsabs.harvard.edu/abs/1997A%26A...326.1135D} {326, 1135}

\bibitem[\protect\citeauthoryear{{Donati} \& {Landstreet}}{{Donati} \&
  {Landstreet}}{2009}]{Donati09}
{Donati} J.,  {Landstreet} J.~D.,  2009, \mn@doi [\araa]
  {10.1146/annurev-astro-082708-101833}, \href
  {http://adsabs.harvard.edu/abs/2009ARA%26A..47..333D} {47, 333}

\bibitem[\protect\citeauthoryear{{Donati}, {Semel}, {Carter}, {Rees}  \&
  {Collier Cameron}}{{Donati} et~al.}{1997}]{Donati97b}
{Donati} J.-F.,  {Semel} M.,  {Carter} B.~D.,  {Rees} D.~E.,   {Collier
  Cameron} A.,  1997, \mnras, \href
  {http://adsabs.harvard.edu/abs/1997MNRAS.291..658D} {291, 658}

\bibitem[\protect\citeauthoryear{{Donati}, {Collier Cameron}  \&
  {Petit}}{{Donati} et~al.}{2003}]{Donati03b}
{Donati} J.-F.,  {Collier Cameron} A.,   {Petit} P.,  2003, \mnras, 345, 1187

\bibitem[\protect\citeauthoryear{{Donati} et~al.,}{{Donati}
  et~al.}{2006}]{Donati06b}
{Donati} J.-F.,  et~al., 2006, \mn@doi [\mnras]
  {10.1111/j.1365-2966.2006.10558.x}, \href
  {http://adsabs.harvard.edu/abs/2006MNRAS.370..629D} {370, 629}

\bibitem[\protect\citeauthoryear{{Donati} et~al.,}{{Donati}
  et~al.}{2020}]{Donati20}
{Donati} J.~F.,  et~al., 2020, \mn@doi [\mnras] {10.1093/mnras/staa2569}, \href
  {https://ui.adsabs.harvard.edu/abs/2020MNRAS.498.5684D} {498, 5684}

\bibitem[\protect\citeauthoryear{{Donati} et~al.,}{{Donati}
  et~al.}{2023a}]{Donati23}
{Donati} J.~F.,  et~al., 2023a, \mn@doi [\mnras] {10.1093/mnras/stad1193},
  \href {https://ui.adsabs.harvard.edu/abs/2023MNRAS.525..455D} {525, 455}

\bibitem[\protect\citeauthoryear{{Donati} et~al.,}{{Donati}
  et~al.}{2023b}]{Donati23b}
{Donati} J.~F.,  et~al., 2023b, \mn@doi [\mnras] {10.1093/mnras/stad2301},
  \href {https://ui.adsabs.harvard.edu/abs/2023MNRAS.525.2015D} {525, 2015}

\bibitem[\protect\citeauthoryear{{Donati} et~al.,}{{Donati}
  et~al.}{2024a}]{Donati24}
{Donati} J.~F.,  et~al., 2024a, \mn@doi [\mnras] {10.1093/mnras/stae675}, \href
  {https://ui.adsabs.harvard.edu/abs/2024MNRAS.tmp..759D} {}

\bibitem[\protect\citeauthoryear{{Donati} et~al.,}{{Donati}
  et~al.}{2024b}]{Donati24b}
{Donati} J.~F.,  et~al., 2024b, \mn@doi [\mnras] {10.1093/mnras/stae1227},
  \href {https://ui.adsabs.harvard.edu/abs/2024MNRAS.531.3256D} {531, 3256}

\bibitem[\protect\citeauthoryear{{Donati}, {Cristofari}, {Carmona}  \&
  {Grankin}}{{Donati} et~al.}{2024c}]{Donati24c}
{Donati} J.~F.,  {Cristofari} P.~I.,  {Carmona} A.,   {Grankin} K.,  2024c,
  \mn@doi [\mnras] {10.1093/mnras/stae2076}, \href
  {https://ui.adsabs.harvard.edu/abs/2024MNRAS.534..231D} {534, 231}

\bibitem[\protect\citeauthoryear{{Dr{\k{a}}{\.z}kowska}
  et~al.,}{{Dr{\k{a}}{\.z}kowska} et~al.}{2023}]{Drazkowska23}
{Dr{\k{a}}{\.z}kowska} J.,  et~al., 2023, in {Inutsuka} S.,  {Aikawa} Y.,
  {Muto} T.,  {Tomida} K.,   {Tamura} M.,  eds,  Astronomical Society of the
  Pacific Conference Series Vol. 534, Protostars and Planets VII. p.~717
  (\mn@eprint {arXiv} {2203.09759}), \mn@doi{10.48550/arXiv.2203.09759}

\bibitem[\protect\citeauthoryear{{Edwards}, {Fischer}, {Kwan}, {Hillenbrand}
  \& {Dupree}}{{Edwards} et~al.}{2003}]{Edwards03}
{Edwards} S.,  {Fischer} W.,  {Kwan} J.,  {Hillenbrand} L.,   {Dupree} A.~K.,
  2003, \mn@doi [\apjl] {10.1086/381077}, \href
  {https://ui.adsabs.harvard.edu/abs/2003ApJ...599L..41E} {599, L41}

\bibitem[\protect\citeauthoryear{{Finociety} \& {Donati}}{{Finociety} \&
  {Donati}}{2022}]{Finociety22}
{Finociety} B.,  {Donati} J.~F.,  2022, \mn@doi [\mnras]
  {10.1093/mnras/stac2682}, \href
  {https://ui.adsabs.harvard.edu/abs/2022MNRAS.516.5887F} {516, 5887}

\bibitem[\protect\citeauthoryear{{Finociety} et~al.,}{{Finociety}
  et~al.}{2023}]{Finociety23b}
{Finociety} B.,  et~al., 2023, \mn@doi [\mnras] {10.1093/mnras/stad3012}, \href
  {https://ui.adsabs.harvard.edu/abs/2023MNRAS.526.4627F} {526, 4627}

\bibitem[\protect\citeauthoryear{{Gaia Collaboration} et~al.,}{{Gaia
  Collaboration} et~al.}{2023}]{Gaia23}
{Gaia Collaboration} et~al., 2023, \mn@doi [\aap]
  {10.1051/0004-6361/202243940}, \href
  {https://ui.adsabs.harvard.edu/abs/2023A&A...674A...1G} {674, A1}

\bibitem[\protect\citeauthoryear{{Gaidos}, {Thanathibodee}, {Hoffman}, {Ong},
  {Hinkle}, {Shappee}  \& {Banzatti}}{{Gaidos} et~al.}{2024}]{Gaidos24}
{Gaidos} E.,  {Thanathibodee} T.,  {Hoffman} A.,  {Ong} J.,  {Hinkle} J.,
  {Shappee} B.~J.,   {Banzatti} A.,  2024, \mn@doi [\apj]
  {10.3847/1538-4357/ad3447}, \href
  {https://ui.adsabs.harvard.edu/abs/2024ApJ...966..167G} {966, 167}

\bibitem[\protect\citeauthoryear{{Gregorio-Hetem} \& {Hetem}}{{Gregorio-Hetem}
  \& {Hetem}}{2002}]{Gregorio02}
{Gregorio-Hetem} J.,  {Hetem} A.,  2002, \mn@doi [\mnras]
  {10.1046/j.1365-8711.2002.05716.x}, \href
  {https://ui.adsabs.harvard.edu/abs/2002MNRAS.336..197G} {336, 197}

\bibitem[\protect\citeauthoryear{{Gregorio-Hetem}, {Lepine}, {Quast}, {Torres}
  \& {de La Reza}}{{Gregorio-Hetem} et~al.}{1992}]{Gregorio92}
{Gregorio-Hetem} J.,  {Lepine} J.~R.~D.,  {Quast} G.~R.,  {Torres} C.~A.~O.,
  {de La Reza} R.,  1992, \mn@doi [\aj] {10.1086/116082}, \href
  {https://ui.adsabs.harvard.edu/abs/1992AJ....103..549G} {103, 549}

\bibitem[\protect\citeauthoryear{{Haffert}, {Bohn}, {de Boer}, {Snellen},
  {Brinchmann}, {Girard}, {Keller}  \& {Bacon}}{{Haffert}
  et~al.}{2019}]{Haffert19}
{Haffert} S.~Y.,  {Bohn} A.~J.,  {de Boer} J.,  {Snellen} I.~A.~G.,
  {Brinchmann} J.,  {Girard} J.~H.,  {Keller} C.~U.,   {Bacon} R.,  2019,
  \mn@doi [Nature Astronomy] {10.1038/s41550-019-0780-5}, \href
  {https://ui.adsabs.harvard.edu/abs/2019NatAs...3..749H} {3, 749}

\bibitem[\protect\citeauthoryear{{Hartmann}, {Herczeg}  \& {Calvet}}{{Hartmann}
  et~al.}{2016}]{Hartmann16}
{Hartmann} L.,  {Herczeg} G.,   {Calvet} N.,  2016, \mn@doi [\araa]
  {10.1146/annurev-astro-081915-023347}, \href
  {http://adsabs.harvard.edu/abs/2016ARA%26A..54..135H} {54, 135}

\bibitem[\protect\citeauthoryear{{Hashimoto} et~al.,}{{Hashimoto}
  et~al.}{2012}]{Hashimoto12}
{Hashimoto} J.,  et~al., 2012, \mn@doi [\apjl] {10.1088/2041-8205/758/1/L19},
  \href {https://ui.adsabs.harvard.edu/abs/2012ApJ...758L..19H} {758, L19}

\bibitem[\protect\citeauthoryear{{Haywood} et~al.,}{{Haywood}
  et~al.}{2014}]{Haywood14}
{Haywood} R.~D.,  et~al., 2014, \mn@doi [\mnras] {10.1093/mnras/stu1320}, \href
  {http://adsabs.harvard.edu/abs/2014MNRAS.443.2517H} {443, 2517}

\bibitem[\protect\citeauthoryear{{Hussain}}{{Hussain}}{2018}]{Hussain18}
{Hussain} G.,  2018, in Take a Closer Look. p.~26,
  \mn@doi{10.5281/zenodo.1488809}

\bibitem[\protect\citeauthoryear{{Isella}, {Benisty}, {Teague}, {Bae},
  {Keppler}, {Facchini}  \& {P{\'e}rez}}{{Isella} et~al.}{2019}]{Isella19}
{Isella} A.,  {Benisty} M.,  {Teague} R.,  {Bae} J.,  {Keppler} M.,  {Facchini}
  S.,   {P{\'e}rez} L.,  2019, \mn@doi [\apjl] {10.3847/2041-8213/ab2a12},
  \href {https://ui.adsabs.harvard.edu/abs/2019ApJ...879L..25I} {879, L25}

\bibitem[\protect\citeauthoryear{Jeffreys}{Jeffreys}{1961}]{Jeffreys61}
Jeffreys H.,  1961, Theory of probability, third edn.
Oxford University Press, Oxford, England

\bibitem[\protect\citeauthoryear{{Johns-Krull}}{{Johns-Krull}}{2007}]{Johns07}
{Johns-Krull} C.~M.,  2007, \mn@doi [\apj] {10.1086/519017}, \href
  {http://adsabs.harvard.edu/abs/2007ApJ...664..975J} {664, 975}

\bibitem[\protect\citeauthoryear{{Keppler} et~al.,}{{Keppler}
  et~al.}{2018}]{Keppler18}
{Keppler} M.,  et~al., 2018, \mn@doi [\aap] {10.1051/0004-6361/201832957},
  \href {https://ui.adsabs.harvard.edu/abs/2018A&A...617A..44K} {617, A44}

\bibitem[\protect\citeauthoryear{{Keppler} et~al.,}{{Keppler}
  et~al.}{2019}]{Keppler19}
{Keppler} M.,  et~al., 2019, \mn@doi [\aap] {10.1051/0004-6361/201935034},
  \href {https://ui.adsabs.harvard.edu/abs/2019A&A...625A.118K} {625, A118}

\bibitem[\protect\citeauthoryear{{Kochanek} et~al.,}{{Kochanek}
  et~al.}{2017}]{Kochanek17}
{Kochanek} C.~S.,  et~al., 2017, \mn@doi [\pasp] {10.1088/1538-3873/aa80d9},
  \href {https://ui.adsabs.harvard.edu/abs/2017PASP..129j4502K} {129, 104502}

\bibitem[\protect\citeauthoryear{{Kuffmeier}}{{Kuffmeier}}{2024}]{Kuffmeier24}
{Kuffmeier} M.,  2024, \mn@doi [arXiv e-prints] {10.48550/arXiv.2406.10901},
  \href {https://ui.adsabs.harvard.edu/abs/2024arXiv240610901K} {p.
  arXiv:2406.10901}

\bibitem[\protect\citeauthoryear{{Kwan}, {Edwards}  \& {Fischer}}{{Kwan}
  et~al.}{2007}]{Kwan07}
{Kwan} J.,  {Edwards} S.,   {Fischer} W.,  2007, \mn@doi [\apj]
  {10.1086/511057}, \href
  {https://ui.adsabs.harvard.edu/abs/2007ApJ...657..897K} {657, 897}

\bibitem[\protect\citeauthoryear{{Landi degl'Innocenti} \& {Landolfi}}{{Landi
  degl'Innocenti} \& {Landolfi}}{2004}]{Landi04}
{Landi degl'Innocenti} E.,  {Landolfi} M.,  2004, {Polarisation in spectral
  lines}.
Dordrecht/Boston/London: Kluwer Academic Publishers

\bibitem[\protect\citeauthoryear{{Langlois} et~al.,}{{Langlois}
  et~al.}{2021}]{Langlois21}
{Langlois} M.,  et~al., 2021, \mn@doi [\aap] {10.1051/0004-6361/202039753},
  \href {https://ui.adsabs.harvard.edu/abs/2021A&A...651A..71L} {651, A71}

\bibitem[\protect\citeauthoryear{{Lau}, {Birnstiel}, {Dr{\k{a}}{\.z}kowska}  \&
  {Stammler}}{{Lau} et~al.}{2024}]{Lau24}
{Lau} T. C.~H.,  {Birnstiel} T.,  {Dr{\k{a}}{\.z}kowska} J.,   {Stammler}
  S.~M.,  2024, \mn@doi [\aap] {10.1051/0004-6361/202450464}, \href
  {https://ui.adsabs.harvard.edu/abs/2024A&A...688A..22L} {688, A22}

\bibitem[\protect\citeauthoryear{{Lehmann} \& {Donati}}{{Lehmann} \&
  {Donati}}{2022}]{Lehmann22}
{Lehmann} L.~T.,  {Donati} J.~F.,  2022, \mn@doi [\mnras]
  {10.1093/mnras/stac1519}, \href
  {https://ui.adsabs.harvard.edu/abs/2022MNRAS.514.2333L} {514, 2333}

\bibitem[\protect\citeauthoryear{{L{\'o}pez-Valdivia}
  et~al.,}{{L{\'o}pez-Valdivia} et~al.}{2021}]{Lopez-Valdivia21}
{L{\'o}pez-Valdivia} R.,  et~al., 2021, \mn@doi [\apj]
  {10.3847/1538-4357/ac1a7b}, \href
  {https://ui.adsabs.harvard.edu/abs/2021ApJ...921...53L} {921, 53}

\bibitem[\protect\citeauthoryear{{Mann} et~al.,}{{Mann} et~al.}{2022}]{Mann22}
{Mann} A.~W.,  et~al., 2022, \mn@doi [\aj] {10.3847/1538-3881/ac511d}, \href
  {https://ui.adsabs.harvard.edu/abs/2022AJ....163..156M} {163, 156}

\bibitem[\protect\citeauthoryear{{Mordasini}, {Alibert}, {Klahr}  \&
  {Henning}}{{Mordasini} et~al.}{2012}]{Mordasini12}
{Mordasini} C.,  {Alibert} Y.,  {Klahr} H.,   {Henning} T.,  2012, \mn@doi
  [\aap] {10.1051/0004-6361/201118457}, \href
  {https://ui.adsabs.harvard.edu/abs/2012A&A...547A.111M} {547, A111}

\bibitem[\protect\citeauthoryear{{Pantolmos}, {Zanni}  \&
  {Bouvier}}{{Pantolmos} et~al.}{2020}]{Pantolmos20}
{Pantolmos} G.,  {Zanni} C.,   {Bouvier} J.,  2020, \mn@doi [\aap]
  {10.1051/0004-6361/202038569}, \href
  {https://ui.adsabs.harvard.edu/abs/2020A&A...643A.129P} {643, A129}

\bibitem[\protect\citeauthoryear{{Pecaut} \& {Mamajek}}{{Pecaut} \&
  {Mamajek}}{2013}]{Pecaut13}
{Pecaut} M.~J.,  {Mamajek} E.~E.,  2013, \mn@doi [\apjs]
  {10.1088/0067-0049/208/1/9}, \href
  {http://adsabs.harvard.edu/abs/2013ApJS..208....9P} {208, 9}

\bibitem[\protect\citeauthoryear{{Press}, {Teukolsky}, {Vetterling}  \&
  {Flannery}}{{Press} et~al.}{1992}]{Press92}
{Press} W.~H.,  {Teukolsky} S.~A.,  {Vetterling} W.~T.,   {Flannery} B.~P.,
  1992, {Numerical recipes in C. The art of scientific computing}

\bibitem[\protect\citeauthoryear{{Pudritz} \& {Ray}}{{Pudritz} \&
  {Ray}}{2019}]{Pudritz19}
{Pudritz} R.~E.,  {Ray} T.~P.,  2019, \mn@doi [Frontiers in Astronomy and Space
  Sciences] {10.3389/fspas.2019.00054}, \href
  {https://ui.adsabs.harvard.edu/abs/2019FrASS...6...54P} {6, 54}

\bibitem[\protect\citeauthoryear{{Rajpaul}, {Aigrain}, {Osborne}, {Reece}  \&
  {Roberts}}{{Rajpaul} et~al.}{2015}]{Rajpaul15}
{Rajpaul} V.,  {Aigrain} S.,  {Osborne} M.~A.,  {Reece} S.,   {Roberts} S.,
  2015, \mn@doi [\mnras] {10.1093/mnras/stv1428}, \href
  {http://adsabs.harvard.edu/abs/2015MNRAS.452.2269R} {452, 2269}

\bibitem[\protect\citeauthoryear{{Romanova}, {Ustyugova}, {Koldoba}  \&
  {Lovelace}}{{Romanova} et~al.}{2004}]{Romanova04}
{Romanova} M.~M.,  {Ustyugova} G.~V.,  {Koldoba} A.~V.,   {Lovelace} R.~V.~E.,
  2004, \mn@doi [\apjl] {10.1086/426586}, \href
  {http://adsabs.harvard.edu/abs/2004ApJ...616L.151R} {616, L151}

\bibitem[\protect\citeauthoryear{{Ryabchikova}, {Piskunov}, {Kurucz},
  {Stempels}, {Heiter}, {Pakhomov}  \& {Barklem}}{{Ryabchikova}
  et~al.}{2015}]{Ryabchikova15}
{Ryabchikova} T.,  {Piskunov} N.,  {Kurucz} R.~L.,  {Stempels} H.~C.,  {Heiter}
  U.,  {Pakhomov} Y.,   {Barklem} P.~S.,  2015, \mn@doi [\physscr]
  {10.1088/0031-8949/90/5/054005}, \href
  {https://ui.adsabs.harvard.edu/abs/2015PhyS...90e4005R} {90, 054005}

\bibitem[\protect\citeauthoryear{{Sanderson}, {Jardine}, {Collier Cameron},
  {Morin}  \& {Donati}}{{Sanderson} et~al.}{2023}]{Sanderson23}
{Sanderson} H.,  {Jardine} M.,  {Collier Cameron} A.,  {Morin} J.,   {Donati}
  J.~F.,  2023, \mn@doi [\mnras] {10.1093/mnras/stac3302}, \href
  {https://ui.adsabs.harvard.edu/abs/2023MNRAS.518.4734S} {518, 4734}

\bibitem[\protect\citeauthoryear{{Skilling} \& {Bryan}}{{Skilling} \&
  {Bryan}}{1984}]{Skilling84}
{Skilling} J.,  {Bryan} R.~K.,  1984, \mn@doi [\mnras]
  {10.1093/mnras/211.1.111}, \href
  {https://ui.adsabs.harvard.edu/abs/1984MNRAS.211..111S} {211, 111}

\bibitem[\protect\citeauthoryear{{Skinner} \& {Audard}}{{Skinner} \&
  {Audard}}{2022}]{Skinner22}
{Skinner} S.~L.,  {Audard} M.,  2022, \mn@doi [\apj]
  {10.3847/1538-4357/ac892f}, \href
  {https://ui.adsabs.harvard.edu/abs/2022ApJ...938..134S} {938, 134}

\bibitem[\protect\citeauthoryear{{Sousa} et~al.,}{{Sousa}
  et~al.}{2023}]{Sousa23}
{Sousa} A.~P.,  et~al., 2023, \mn@doi [\aap] {10.1051/0004-6361/202244720},
  \href {https://ui.adsabs.harvard.edu/abs/2023A&A...670A.142S} {670, A142}

\bibitem[\protect\citeauthoryear{{Thanathibodee}, {Calvet}, {Bae}, {Muzerolle}
  \& {Hern{\'a}ndez}}{{Thanathibodee} et~al.}{2019}]{Thanathibodee19}
{Thanathibodee} T.,  {Calvet} N.,  {Bae} J.,  {Muzerolle} J.,   {Hern{\'a}ndez}
  R.~F.,  2019, \mn@doi [\apj] {10.3847/1538-4357/ab44c1}, \href
  {https://ui.adsabs.harvard.edu/abs/2019ApJ...885...94T} {885, 94}

\bibitem[\protect\citeauthoryear{{Thanathibodee} et~al.,}{{Thanathibodee}
  et~al.}{2020}]{Thanathibodee20}
{Thanathibodee} T.,  et~al., 2020, \mn@doi [\apj] {10.3847/1538-4357/ab77c1},
  \href {https://ui.adsabs.harvard.edu/abs/2020ApJ...892...81T} {892, 81}

\bibitem[\protect\citeauthoryear{{Thanathibodee}, {Calvet}, {Hern{\'a}ndez},
  {Mauc{\'o}}  \& {Brice{\~n}o}}{{Thanathibodee}
  et~al.}{2022}]{Thanathibodee22}
{Thanathibodee} T.,  {Calvet} N.,  {Hern{\'a}ndez} J.,  {Mauc{\'o}} K.,
  {Brice{\~n}o} C.,  2022, \mn@doi [\aj] {10.3847/1538-3881/ac3ee6}, \href
  {https://ui.adsabs.harvard.edu/abs/2022AJ....163...74T} {163, 74}

\bibitem[\protect\citeauthoryear{{Tsukamoto} et~al.,}{{Tsukamoto}
  et~al.}{2023}]{Tsukamoto23}
{Tsukamoto} Y.,  et~al., 2023, in {Inutsuka} S.,  {Aikawa} Y.,  {Muto} T.,
  {Tomida} K.,   {Tamura} M.,  eds,  Astronomical Society of the Pacific
  Conference Series Vol. 534, Protostars and Planets VII. p.~317 (\mn@eprint
  {arXiv} {2209.13765}), \mn@doi{10.48550/arXiv.2209.13765}

\bibitem[\protect\citeauthoryear{{Ustyugova}, {Koldoba}, {Romanova}  \&
  {Lovelace}}{{Ustyugova} et~al.}{2006}]{Ustyugova06}
{Ustyugova} G.~V.,  {Koldoba} A.~V.,  {Romanova} M.~M.,   {Lovelace} R.~V.~E.,
  2006, \mn@doi [\apj] {10.1086/503379}, \href
  {http://adsabs.harvard.edu/abs/2006ApJ...646..304U} {646, 304}

\bibitem[\protect\citeauthoryear{{Xie}, {Haffert}, {de Boer}, {Kenworthy},
  {Brinchmann}, {Girard}, {Snellen}  \& {Keller}}{{Xie} et~al.}{2020}]{Xie20}
{Xie} C.,  {Haffert} S.~Y.,  {de Boer} J.,  {Kenworthy} M.~A.,  {Brinchmann}
  J.,  {Girard} J.,  {Snellen} I.~A.~G.,   {Keller} C.~U.,  2020, \mn@doi
  [\aap] {10.1051/0004-6361/202038242}, \href
  {https://ui.adsabs.harvard.edu/abs/2020A&A...644A.149X} {644, A149}

\bibitem[\protect\citeauthoryear{{Zanni} \& {Ferreira}}{{Zanni} \&
  {Ferreira}}{2013}]{Zanni13}
{Zanni} C.,  {Ferreira} J.,  2013, \mn@doi [\aap]
  {10.1051/0004-6361/201220168}, \href
  {http://adsabs.harvard.edu/abs/2013A%26A...550A..99Z} {550, A99}

\makeatother
\end{thebibliography}
\bibliographystyle{mnras}

\appendix

\section{ASAS-SN light curves and stacked periodograms}
\label{sec:appA}

\begin{figure*}
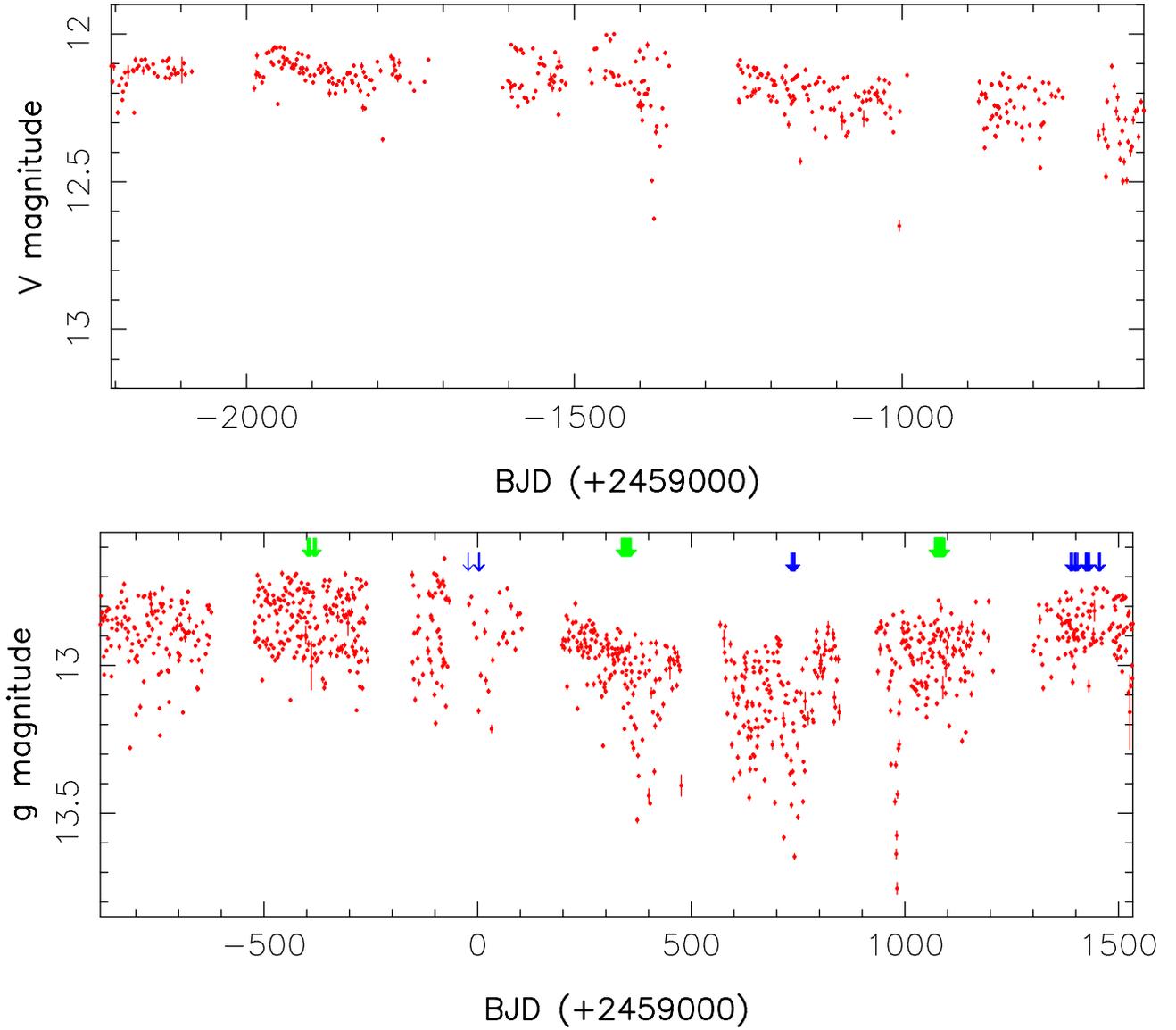

\centerline{\includegraphics[scale=0.7,angle=-90]{fig/pds70-lc4.ps}\vspace{4mm}}
\centerline{\includegraphics[scale=0.7,angle=-90]{fig/pds70-lc3.ps}}
\caption[]{ASAS-SN  $V$-band (2014 May to 2018 September, top panel) and $g$-band (2017 December to 2024 August, bottom panel) light curves of PDS~70, binned down to one point 
per night (red dots with error bars).  On the second panel, we also depict the epochs of our SPIRou spectra (blue arrows) and of the TESS observations (sectors 11, 38 and 65, 
green arrows).  Both panels have the same dynamic range to emphasize the temporal variations of the photometric fluctuations.   } 
\label{fig:lc2}
\end{figure*}

\begin{figure*}
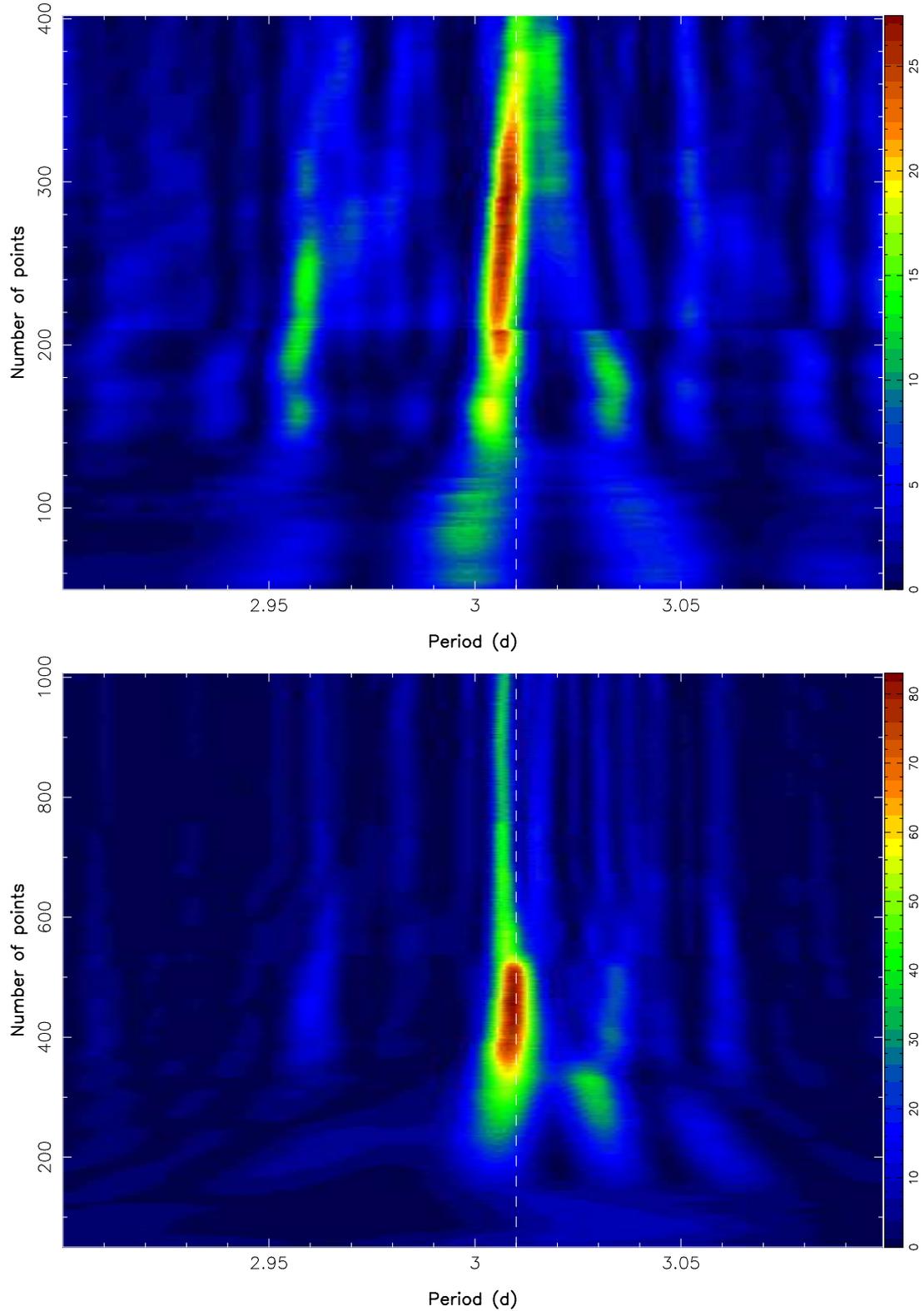

\centerline{\includegraphics[scale=0.5,angle=-90]{fig/pds70-lc1.ps}}
\centerline{\includegraphics[scale=0.5,angle=-90]{fig/pds70-lc2.ps}}
\caption[]{Stacked periodograms of the ASAS-SN $V$-band (top panel) and $g$-band (bottom panel) light curves of PDS~70 (see Fig.~\ref{fig:lc2}).   
In both plots, each horizontal line corresponds to a color-coded periodogram, computed on an increasing number of points (starting from the first one).  
The color scale (different in both panels) depicts the logarithmic power of each periodogram.  A signal at a period of about 3~d is detected in both cases, 
its exact position and strength varying with time and weakening towards the end.  The vertical dashed line depicts the rotation period as derived from our study.  }
\label{fig:lc}
\end{figure*}

We show in Fig.~\ref{fig:lc2} the ASAS-SN $V$-band (2014 May to 2018 September) and $g$-band (2017 December to 2024 August) light curves of PDS~70 (binned down to one point per night), 
along with the epochs of our SPIRou spectra and TESS observations.  We also show in Fig.~\ref{fig:lc} the stacked periodograms of both light curves.  

\section{Observation log}
\label{sec:appB}

Table~\ref{tab:log} gives the full log and associated \Bl\ and RV measurements at each observing epoch from our SPIRou spectra.

\begin{table*}
\small
\caption[]{Observing log of our SPIRou observations of PDS~70 in seasons 2020, 2022 and 2024.  All exposures consist of 4 sub-exposures of equal length, 
except on 2022 June 12, when bad weather only allowed for a pair of sub-exposures to be collected.
For each visit, we list the barycentric Julian date BJD, the UT date, the rotation cycle c and phase $\phi$ (computed as indicated in Sec.~\ref{sec:obs}),
the total observing time t$_{\rm exp}$, the peak SNR in the spectrum (in the $H$ band) per 2.3~\kms\ pixel, the noise level in the LSD Stokes $V$ profile,
the estimated \Bl\ with error bars, the nightly averaged RVs, {\emr BISs and FWHMs} and corresponding error bars derived from Stokes $I$ LSD profiles of atomic lines, 
the RVs and corresponding error bars of CO bandhead lines, 
and the veiling in the $JH$ and in the $K$ bands $r_{JH}$ and $r_K$ measured from LSD profiles of atomic lines and CO lines respectively.  } 
\scalebox{0.95}{\hspace{-4mm}
\begin{tabular}{cccccccccccc}
\hline
BJD        & UT date & c / $\phi$ & t$_{\rm exp}$ & SNR & $\sigma_V$            & \Bl\   &  RV    & BIS    & FWHM   &   RV (CO) &  $r_{JH}$ / $r_K$ \\
(2459000+) &         &            &   (s)        & ($H$) & ($10^{-4} I_c$)       & (G)   & (\kms) & (\kms) & (\kms) & (\kms)    &                   \\
\hline
-22.0907349 & 08 May 2020 & 0 / 0.003 & 2808.2 & 186 & 2.93 & -37$\pm$19 & 5.79$\pm$0.17 & 1.93$\pm$0.21 & 32.3$\pm$0.5 & 6.06$\pm$0.55 & 0.18 / 0.56 \\
1.8378984 & 01 Jun 2020 & 7 / 0.953 & 2808.2 & 183 & 3.50 & -43$\pm$21 & 6.48$\pm$0.18 & 0.91$\pm$0.22 & 31.2$\pm$0.5 & 6.58$\pm$0.23 & 0.11 / 0.53 \\
3.8672408 & 03 Jun 2020 & 8 / 0.627 & 2808.2 & 230 & 2.23 & 93$\pm$13 & 6.98$\pm$0.17 & -0.70$\pm$0.22 & 31.3$\pm$0.5 & 6.51$\pm$0.24 & 0.08 / 0.46 \\
4.8240857 & 04 Jun 2020 & 8 / 0.945 & 2808.2 & 224 & 2.25 & -31$\pm$14 & 5.93$\pm$0.16 & 0.92$\pm$0.20 & 31.7$\pm$0.5 & 6.59$\pm$0.20 & 0.13 / 0.43 \\
\hline
732.8321354 & 02 Jun 2022 & 250 / 0.808 & 2808.2 & 215 & 2.40 & 75$\pm$15 & 6.44$\pm$0.17 & 0.04$\pm$0.21 & 28.8$\pm$0.5 & 6.79$\pm$0.22 & 0.13 / 0.57 \\
733.8368940 & 03 Jun 2022 & 251 / 0.142 & 2808.2 & 235 & 2.18 & 64$\pm$13 & 6.70$\pm$0.18 & -0.95$\pm$0.22 & 30.5$\pm$0.5 & 6.89$\pm$0.22 & 0.10 / 0.58 \\
735.8297289 & 05 Jun 2022 & 251 / 0.804 & 2808.2 & 241 & 2.10 & 55$\pm$12 & 6.26$\pm$0.16 & 0.19$\pm$0.20 & 29.0$\pm$0.5 & 6.64$\pm$0.18 & 0.10 / 0.72 \\
736.8314022 & 06 Jun 2022 & 252 / 0.137 & 2808.2 & 233 & 2.17 & 55$\pm$13 & 6.62$\pm$0.17 & -0.03$\pm$0.21 & 31.1$\pm$0.5 & 6.88$\pm$0.20 & 0.10 / 0.50 \\
737.8318903 & 07 Jun 2022 & 252 / 0.469 & 2808.2 & 215 & 2.93 & 12$\pm$17 & 6.38$\pm$0.16 & 0.86$\pm$0.20 & 30.1$\pm$0.5 & 6.44$\pm$0.20 & 0.11 / 0.54 \\
739.7949216 & 09 Jun 2022 & 253 / 0.121 & 2808.2 & 242 & 2.04 & 69$\pm$12 & 6.61$\pm$0.15 & -0.58$\pm$0.19 & 31.4$\pm$0.5 & 7.08$\pm$0.26 & 0.10 / 0.35 \\
740.8243867 & 10 Jun 2022 & 253 / 0.463 & 2808.2 & 230 & 2.14 & 29$\pm$12 & 5.89$\pm$0.15 & 1.12$\pm$0.19 & 30.8$\pm$0.5 & 6.24$\pm$0.19 & 0.05 / 0.44 \\
741.8079906 & 11 Jun 2022 & 253 / 0.790 & 2808.2 & 224 & 2.21 & 74$\pm$13 & 5.64$\pm$0.15 & 0.93$\pm$0.19 & 29.5$\pm$0.5 & 6.02$\pm$0.19 & 0.13 / 0.98 \\
742.7851079 & 12 Jun 2022 & 254 / 0.115 & 1404.1 & 110 & 4.63 & 86$\pm$30 & 6.93$\pm$0.17 & -0.31$\pm$0.21 & 31.1$\pm$0.5 & 6.66$\pm$0.28 & 0.18 / 0.61 \\
\hline
1387.0325927 & 17 Mar 2024 & 468 / 0.150 & 2808.2 & 206 & 2.30 & 176$\pm$13 & 6.81$\pm$0.16 & -0.35$\pm$0.20 & 31.7$\pm$0.5 & 7.05$\pm$0.19 & 0.01 / 0.60 \\
1388.0312781 & 18 Mar 2024 & 468 / 0.482 & 2808.2 & 215 & 2.22 & 71$\pm$13 & 6.00$\pm$0.16 & -0.06$\pm$0.20 & 31.4$\pm$0.5 & 6.30$\pm$0.18 & 0.06 / 0.61 \\
1389.0227861 & 19 Mar 2024 & 468 / 0.812 & 2808.2 & 238 & 1.95 & -104$\pm$11 & 5.85$\pm$0.15 & 0.88$\pm$0.19 & 30.7$\pm$0.5 & 6.36$\pm$0.18 & 0.01 / 0.72 \\
1391.0457659 & 21 Mar 2024 & 469 / 0.484 & 2808.2 & 187 & 2.61 & 86$\pm$15 & 6.11$\pm$0.16 & 0.88$\pm$0.20 & 32.0$\pm$0.5 & 6.62$\pm$0.25 & 0.09 / 0.32 \\
1391.9946767 & 22 Mar 2024 & 469 / 0.799 & 2808.2 & 195 & 2.48 & -103$\pm$14 & 5.75$\pm$0.16 & 1.08$\pm$0.20 & 30.7$\pm$0.5 & 6.57$\pm$0.26 & 0.04 / 0.31 \\
1393.0092961 & 23 Mar 2024 & 470 / 0.136 & 2808.2 & 190 & 2.53 & 143$\pm$14 & 6.36$\pm$0.16 & 0.13$\pm$0.19 & 32.1$\pm$0.5 & 6.64$\pm$0.21 & 0.03 / 0.37 \\
1397.0023396 & 27 Mar 2024 & 471 / 0.463 & 2808.2 & 223 & 2.12 & 64$\pm$13 & 6.33$\pm$0.18 & 0.13$\pm$0.22 & 31.5$\pm$0.5 & 6.81$\pm$0.22 & 0.10 / 1.02 \\
1399.0130591 & 29 Mar 2024 & 472 / 0.131 & 2808.2 & 208 & 2.32 & 113$\pm$13 & 6.80$\pm$0.17 & -0.21$\pm$0.21 & 32.3$\pm$0.5 & 7.37$\pm$0.20 & 0.03 / 0.46 \\
1399.9973197 & 30 Mar 2024 & 472 / 0.458 & 2808.2 & 236 & 1.99 & 63$\pm$10 & 6.59$\pm$0.17 & 0.63$\pm$0.21 & 33.2$\pm$0.5 & 6.70$\pm$0.19 & -0.06 / 0.31 \\
1401.0273243 & 31 Mar 2024 & 472 / 0.800 & 2808.2 & 195 & 2.50 & -116$\pm$13 & 5.78$\pm$0.16 & 1.31$\pm$0.20 & 30.8$\pm$0.5 & 6.49$\pm$0.20 & -0.03 / 0.31 \\
1401.9923358 & 01 Apr 2024 & 473 / 0.120 & 2808.2 & 232 & 1.99 & 128$\pm$10 & 6.86$\pm$0.18 & 0.01$\pm$0.22 & 32.7$\pm$0.5 & 7.09$\pm$0.20 & -0.09 / 0.23 \\
1405.0097796 & 04 Apr 2024 & 474 / 0.123 & 2808.2 & 165 & 3.15 & 135$\pm$18 & 7.28$\pm$0.17 & -0.45$\pm$0.21 & 30.2$\pm$0.5 & 7.48$\pm$0.24 & 0.07 / 0.43 \\
1421.9643985 & 21 Apr 2024 & 479 / 0.756 & 2808.2 & 216 & 2.24 & -109$\pm$12 & 5.55$\pm$0.16 & 0.92$\pm$0.20 & 31.1$\pm$0.5 & 6.46$\pm$0.20 & 0.03 / 0.43 \\
1422.9497230 & 22 Apr 2024 & 480 / 0.083 & 2808.2 & 241 & 2.00 & 91$\pm$11 & 6.72$\pm$0.16 & -0.55$\pm$0.21 & 30.4$\pm$0.5 & 7.09$\pm$0.20 & 0.03 / 0.46 \\
1423.9116919 & 23 Apr 2024 & 480 / 0.403 & 2808.2 & 230 & 2.10 & 124$\pm$11 & 6.56$\pm$0.16 & -0.39$\pm$0.20 & 32.3$\pm$0.5 & 6.70$\pm$0.20 & 0.01 / 0.41 \\
1424.8707088 & 24 Apr 2024 & 480 / 0.721 & 2808.2 & 245 & 1.90 & -68$\pm$10 & 5.80$\pm$0.16 & 1.39$\pm$0.20 & 31.5$\pm$0.5 & 6.63$\pm$0.21 & 0.00 / 0.38 \\
1425.9189887 & 25 Apr 2024 & 481 / 0.069 & 2808.2 & 206 & 2.37 & 78$\pm$13 & 6.33$\pm$0.16 & -0.26$\pm$0.20 & 30.7$\pm$0.5 & 6.81$\pm$0.23 & 0.06 / 0.43 \\
1426.9430337 & 26 Apr 2024 & 481 / 0.410 & 2808.2 & 217 & 2.19 & 113$\pm$12 & 6.79$\pm$0.16 & -0.21$\pm$0.20 & 32.0$\pm$0.5 & 6.79$\pm$0.27 & 0.05 / 0.30 \\
1427.8992940 & 27 Apr 2024 & 481 / 0.727 & 2808.2 & 243 & 1.97 & -64$\pm$11 & 5.58$\pm$0.16 & 1.15$\pm$0.20 & 30.8$\pm$0.5 & 6.87$\pm$0.21 & 0.02 / 0.34 \\
1429.9103948 & 29 Apr 2024 & 482 / 0.395 & 2808.2 & 120 & 4.77 & 172$\pm$29 & 6.04$\pm$0.18 & 0.16$\pm$0.22 & 31.1$\pm$0.5 & 6.87$\pm$0.31 & 0.14 / 0.51 \\
1431.9180016 & 01 May 2024 & 483 / 0.062 & 2808.2 & 234 & 2.09 & 68$\pm$11 & 6.75$\pm$0.16 & -0.24$\pm$0.20 & 31.2$\pm$0.5 & 7.29$\pm$0.36 & -0.01 / 0.20 \\
1432.9104947 & 02 May 2024 & 483 / 0.392 & 2808.2 & 227 & 2.17 & 145$\pm$12 & 6.53$\pm$0.17 & -0.59$\pm$0.21 & 31.6$\pm$0.5 & 7.20$\pm$0.30 & 0.03 / 0.31 \\
1452.9013731 & 22 May 2024 & 490 / 0.034 & 2808.2 & 232 & 2.17 & 39$\pm$12 & 6.81$\pm$0.17 & -0.62$\pm$0.21 & 30.4$\pm$0.5 & 7.18$\pm$0.22 & 0.02 / 0.29 \\
1454.8762118 & 24 May 2024 & 490 / 0.690 & 2808.2 & 187 & 2.73 & -45$\pm$16 & 5.29$\pm$0.17 & 0.98$\pm$0.22 & 30.9$\pm$0.5 & 6.47$\pm$0.25 & 0.08 / 0.52 \\
1455.8555552 & 25 May 2024 & 491 / 0.015 & 2808.2 & 183 & 3.47 & 60$\pm$19 & 6.50$\pm$0.17 & -0.48$\pm$0.21 & 30.9$\pm$0.5 & 7.20$\pm$0.25 & 0.03 / 0.62 \\
1456.8256793 & 26 May 2024 & 491 / 0.337 & 2808.2 & 133 & 4.26 & 94$\pm$26 & 7.08$\pm$0.18 & -1.26$\pm$0.23 & 32.0$\pm$0.5 & 7.56$\pm$0.30 & 0.14 / 1.17 \\
1457.8493597 & 27 May 2024 & 491 / 0.678 & 2808.2 & 241 & 2.06 & -24$\pm$11 & 5.59$\pm$0.17 & 1.18$\pm$0.21 & 31.5$\pm$0.5 & 6.77$\pm$0.21 & -0.02 / 0.34 \\
\hline
\end{tabular}}
\label{tab:log}
\end{table*}

\section{RV periodograms} 
\label{sec:appC}

\begin{figure*}
\centerline{\includegraphics[scale=0.35,angle=-90]{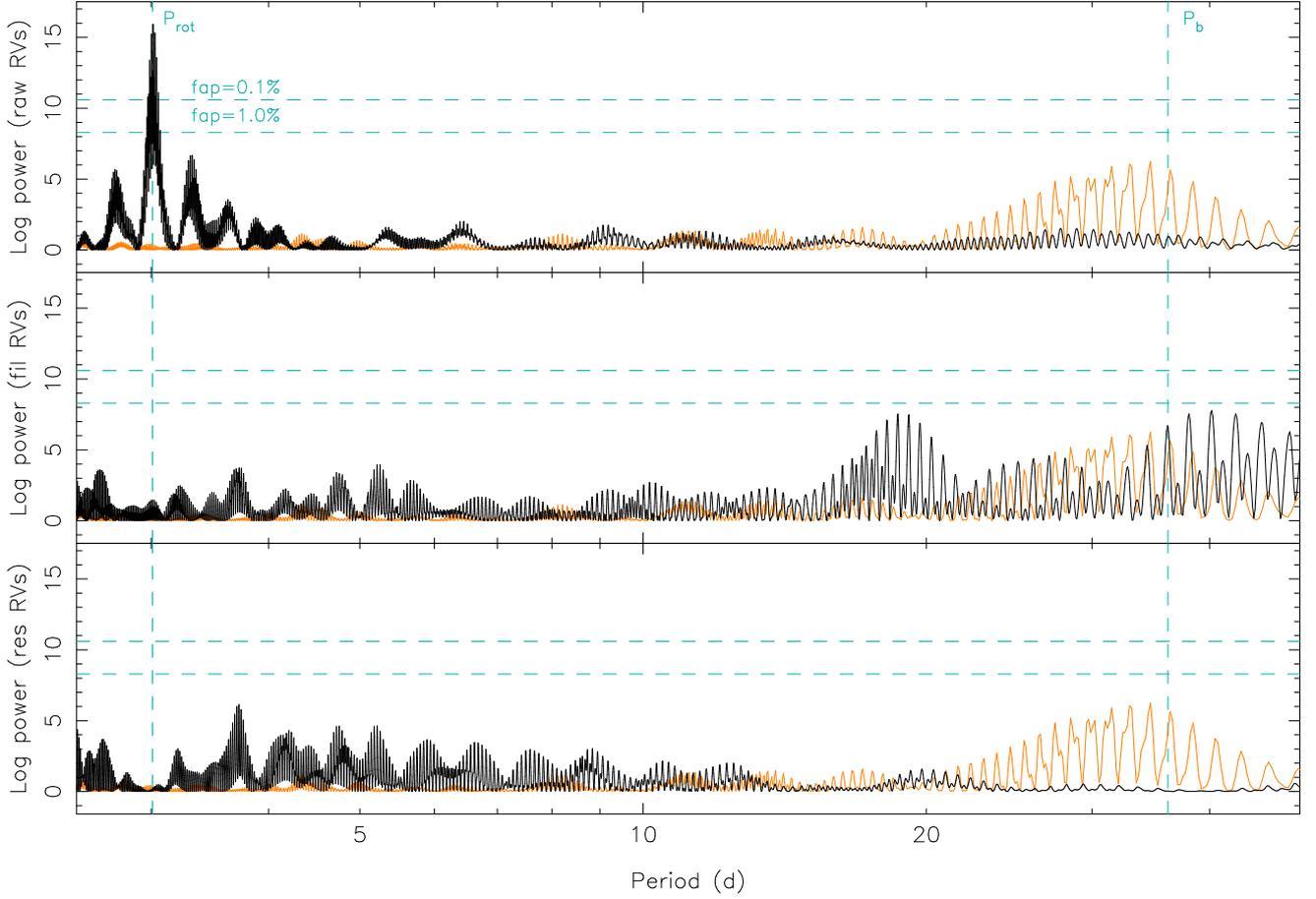}} 
\caption[]{Periodograms of the raw RVs (top plot), {\emr filtered RVs (i.e., raw RVs subtracted from the adjusted GP, middle plot) and residual RVs (i.e., filtered 
RVs subtracted from the additional sinusoidal pattern, bottom plot) when including a sinusoidal fit at period of $P_b$ in addition to the GPR modeling of the raw RVs} 
in the MCMC process.  The cyan vertical dashed lines trace the derived \Prot\ and $P_b$ whereas the horizontal dashed lines indicate the 1 and 0.1~per cent FAP levels in 
the periodograms of our RV data.  The orange curve depicts the window function.  
Excess power is visible in the periodogram of the filtered RVs around 20~d and 40~d 
that is best modelled by a sinusoidal RV signal at $P_b=36.16$~d (see Sec.~\ref{sec:rvs}). } 
\label{fig:per}
\end{figure*}

We show in Fig.~\ref{fig:per} the periodograms of raw RVs, filtered RVs (i.e., raw RVs subtracted from the adjusted GP, see text) and residual RVs (i.e., filtered 
RVs subtracted from the additional sinusoidal pattern, see text) of PDS~70, resulting from the analysis described in Sec.~\ref{sec:rvs}.  

\bsp    
\label{lastpage}
\end{document}